%% file: PMR.tex
\documentclass[10pt, conference, letterpaper]{IEEEtran}

\usepackage{amsmath}
\usepackage{cases}
\usepackage{graphicx}

\usepackage{cite}
\usepackage{times}

\usepackage{graphicx}
\usepackage{graphics}
\usepackage{subfigure}
\usepackage{epsfig}
\usepackage{latexsym}
\usepackage{amsfonts}
\usepackage{amssymb}
\usepackage{paralist}
\usepackage{xspace}
\usepackage{mathrsfs}
\usepackage{amssymb}
\usepackage{color}
\usepackage[ruled]{algorithm}
\usepackage[noend]{algorithmic}
\usepackage{mathtools}

\usepackage{amssymb}
\usepackage{url,epsfig,array}
\usepackage{tikz}

\usepackage[T1]{fontenc}
\usepackage[utf8]{inputenc}
\usepackage{lmodern}

\allowdisplaybreaks 
\def\ie{\textit{i.e.}\xspace}

\def\eg{\textit{e.g.}\xspace}

\newcommand{\CUTXY}[1]{{}}
\newcommand{\CUTTH}[1]{{}}
\newcommand{\KA}{KA\xspace}

\newcommand{\abe}{\texttt{ABE.E}}

\newcommand{\hed}{\texttt{HE.D}}
\newcommand{\enh}{\texttt{HE.E}}

\newcommand{\fpr}{\texttt{f}}

\newcommand{\dis}{\mathsf{d}}
\newcommand{\sdis}{\mathsf{D}}

\newcommand{\vV}{\mathbf{V}}
\newcommand{\vv}{\mathbf{v}}
\newcommand{\vr}{\mathbf{r}}
\newcommand{\vq}{\mathbf{q}}
\newcommand{\vC}{\mathbf{C}}
\newcommand{\vl}{\mathbf{l}}
\newcommand{\vf}{\mathbf{f}}

\newcommand{\vx}{\mathbf{x}}
\newcommand{\vy}{\mathbf{y}}
\newcommand{\fx}{\mathbf{X}}
\newcommand{\fy}{\mathbf{Y}}
\newcommand{\fd}{\mathbf{D}}

\renewcommand{\paragraph}[1]{\smallskip \noindent {\textbf{#1}}}

\newtheorem{definition}{Definition}

\begin{document}

\title{Cloud-based Privacy Preserving Image Storage, Sharing and Search}


\author{\authorblockN{Lan Zhang\authorrefmark{1},
Taeho Jung \authorrefmark{2},
Puchun Feng \authorrefmark{1},
Xiang-Yang Li\authorrefmark{2},
Yunhao Liu\authorrefmark{1}}
\authorblockA{\authorrefmark{1} School of Software, Tsinghua University}
\authorblockA{\authorrefmark{2} Department of Computer Science, Illinois Institute of Technology}
}

\maketitle

\begin{abstract}
\input{abstract.tex}
\end{abstract}

\section{Introduction}
\label{sec:introduction}
\input{intro.tex}

\section{Background on Large-scale Image Search}
\label{sec:prelim}
\input{prelim.tex}

\section{Multi-level Homomorphic Encryption}
\label{section:HE}
\input{he.tex}

\section{System Overview}
\label{sec:model}
\input{model.tex}

\section{Basic System Design}
\label{sec:design}
\input{approach.tex}

\section{System Refinements}
\label{sec:refine}
\input{system.tex}

\section{Security Analysis}
\label{sec:ana}

\input{ana.tex}

\section{Implementation and Evaluation}
\label{sec:eva}
\input{eva.tex}

\vspace{-0.08in}
\section{Related work}
\label{sec:review}
\input{related.tex}

\vspace{-0.08in}
\section{Conclusion}
\label{sec:conclusion}
\input{conclusion.tex}


\vspace{-0.08in}
{
\bibliographystyle{IEEEtran}
\bibliography{ref}
}

\end{document}

%% file: abstract.tex
High-resolution cameras produce huge volume of high quality images everyday.
It is extremely challenging to store, share and especially search those huge images,
for which increasing number of cloud services are presented to support such functionalities.
However, images tend to contain rich sensitive information (\eg, people, location and event),
 and people's privacy concerns hinder their readily participation into the services
 provided by untrusted third parties.

In this work,
 we introduce PIC: a Privacy-preserving large-scale Image search system on Cloud.
Our system enables efficient yet secure content-based image search with fine-grained access control,
 and it also provides privacy-preserving image storage and sharing among users.
Users can specify who can/cannot search on their images when using the system,
 and they can search on others' images if they satisfy the condition specified by the image owners.
Majority of the computationally intensive jobs are outsourced to the cloud side,
 and users only need to submit the query and receive the result throughout the entire image search.
Specially, to deal with massive images,
 we design our system suitable for distributed and parallel computation
 and introduce several optimizations to further expedite the search process.

We implement a prototype of PIC including both cloud side and client side.
The cloud side is a cluster of computers with distributed file system (Hadoop HDFS)
and MapReduce architecture (Hadoop MapReduce).
The client side is built for both Windows OS laptops and Android phones.
We evaluate the prototype system with large sets of real-life photos.
Our security analysis and evaluation results show that PIC successfully
 protect the image privacy at a low cost of computation and communication.

%% file: intro.tex
As on-board cameras get more and more popular,
 numerous high-resolution photos are generated everyday,
 which makes storing, sharing and especially searching
 large-scale images become challenging issues.
There are an increasing number of service providers supporting cloud-based image services,
 \eg, Amazon Cloud Drive, Apple iCloud, Cloudinary, Flicker and Google.
Content-based image search is a core functionality for a variety of image applications,
  such as personal image management, criminal investigation using crowd-sourced photos (\eg, the Boston Marathon investigation \cite{Boston})
  and medical image study and diagnosis \cite{korn1996fast,tagare1997medical}.
It draws many attentions from both industry (\eg, Google) and academics (\eg, \cite{jegou2007contextual, jegou2010improving}).
However, there are a lot of private information in images and the cloud could be untrustworthy \cite{ion2011home},
 which hinder the wide adoption of those useful image services.
While leveraging the power of cloud and crowdsourcing, image services should be provided in a privacy-preserving way.
Besides, the image owner also requires the right to determine who is valid to search and access his images.

\begin{figure}[h]
\centering
\subfigure[Compare original image and reconstructed image using SIFT feature vector.\cite{weinzaepfel2011reconstructing}]{\includegraphics[width=0.7 \linewidth, height=0.88in]{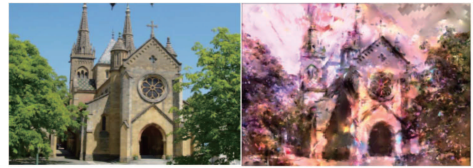}}\\
\subfigure[Feature vector detection and matching results between original image and reconstructed image.]{\includegraphics[width=0.7 \linewidth, height=0.88in]{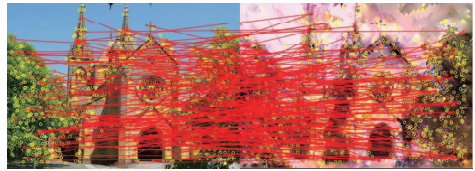}}
 \caption{Comparison between original image and images reconstructed from feature vectors.}
 \label{fig:feature}
\label{reconstruct}
\end{figure}

The state-of-the-art image search systems typically use feature descriptors of images
 to measure their content similarity, \eg, SIFT \cite{lowe2004distinctive}.
Feature descriptors are sets of high-dimension feature vectors, \eg, 128-dimension for SIFT \cite{lowe2004distinctive},
 extracted from interest points of the images (Fig.~\ref{fig:feature}).
Those feature descriptors are distinctive and show good accuracy
 for image search.
One image is usually described by hundreds of feature vectors,
then the millions of photos uploaded to the cloud imply that billions of feature vectors need to be processed.
Therefore, it is necessary to introduce some optimization techniques such as indexing or distributed computing
 to accelerate the search process.
Using feature vectors, many indexing mechanisms are designed to facilitate large-scale image search, \eg, \cite{jegou2009recent}, \cite{jegou2010improving}
and \cite{gudmundsson2010large}.
But most of them did not support image privacy protection or assumed that the feature vectors do not reveal useful information about the original images.
However, recent research shows that
 an image can be approximately reconstructed based on the output of a blackbox
 feature descriptor software such as those classically used for image indexing \cite{weinzaepfel2011reconstructing,daneshi2011image}.
As presented in Fig.~\ref{fig:feature},
 the image reconstructed using SIFT feature vectors appears quite similar as the original image,
 and shows a good matching result with the original one.
Even quantized binary feature vectors expose the content of an image to some extent \cite{d2012beyond}.
Those methods are entirely automatic and much better reconstruction can be achieved with user interaction,
 which arouse great concerns on the image content privacy in the image indexing and search systems.

Recently, some facts show that people pay more and more attention to the privacy of their images,
 and the increasing worry about privacy may be the key point in the future cloud-based image services.
For example, a smartphone app KeepSafe \cite{Keepsafe} hides images and videos
  from unauthorized users by simply encrypting image folders with a PIN code,
  and it owns 13 million users and gained an investment of 2 millions dollars.
Furthermore, the face search functionality of Facebook has been abandoned for two years due to
 the privacy concern from users and governments \cite{facebook_off}.

As a result,
 massive image need efficient yet private content-based search functions urgently.
Existing large-scale image indexing and search systems usually ignore the privacy issue and do not support privacy protection mechanisms.
Some systems tag images with keywords and metadata, and search on encrypted keywords, \eg, \cite{song2000practical,golle2004secure, lu2011privacy, cong2012keyword}.
But keyword-based search methods cannot be adopted for content-based image search,
 because they only search the occurrences of encoded tags in an exact match manner
 and cannot measure the distance between high-dimension feature vectors.
Moreover, they usually reveal the search results to the cloud.
To achieve content-based image search, we need to measure the distance between pairs of feature vectors as well as to protect the vectors and search results.
This problem is especially challenging when we require no interaction with the image owner during the whole search process and the image data are extremely large (e.g., with millions of feature vectors).
Traditional secure multi-party computation (SMC) \cite{yao1982protocols,lindell2004proof}, verifiable computation \cite{parno2013pinocchio}
 or simple homomorphic encryption\cite{erkin2009privacy,sadeghi2010efficient} support privacy-preserving vector distance computation.
But they cannot be the solution to the large-scale image search problem.
On one hand, the garbled circuit's size is exponentially greater than the size of the input, and the input size in an image search is often very large (\eg, thousands of 128-dimensional feature vectors of real values), so the communication and computation overhead is beyond practicality.
On the other hand, a simple homomorphic method will allow the decrypter of the final result to also decrypt the individual ciphertexts.
Moreover, it requires rounds of interactions between the querier, data owner and cloud server,
 which means the image owners need to always stay online to react to image queries, which is not practical either.
Facing these challenges, we employ the multi-level homomorphic encryption (Xiao \textit{et al} \cite{xiao2012efficient}) as a building block
 to design an efficient non-interactive image search system.

The main contributions of this paper can be summarized as follows:
\begin{enumerate}
\item We propose a novel system (PIC) and techniques to enable private content-based image search upon large-scale encrypted images with untrusted servers.
 Our system outsources the majority of the search job to the cloud side, but neither the query result nor the query itself is revealed to the cloud.
 What's more, during the search, no interaction is required between the data owner and the querier or the cloud.

\item We design our private search compatible with the standard image search to guarantee the search accuracy.
  We also make it suitable for distributed and parallel computation to enable efficient large-scale image search.
  Several optimizations are introduced to further expedite the search process.

\item  We implemented a prototype including both cloud side and client side.
    The cloud is a cluster of computers with distributed file system (Hadoop HDFS) and MapReduce architecture (Hadoop MapReduce).
    We evaluate our system using large sets of highly diverse real-life photos.
     The security analysis shows that PIC successfully protect the image privacy,
    and the evaluation shows the efficiency of our system, which is capable to be used in a wide range of platforms including resource-bounded ones.
\end{enumerate}

\paragraph{Roadmap.}The rest of this paper is organized as follows. Section \ref{sec:prelim}
reviews the image search model and Section \ref{section:HE} introduces the building blocks of PIC.
Section \ref{sec:model} gives an overview of our system framework.
Section \ref{sec:design} presents the details of our basic design,
 then we discuss the bottleneck of system performance and refine system design in Section \ref{sec:refine}.
We show the security of PIC in Section \ref{sec:ana}.
To evaluate the practicality, we present comprehensive experiments in Section \ref{sec:eva}.
Section \ref{sec:review} discusses the related work, and Section \ref{sec:conclusion} concludes
this paper.

%% file: prelim.tex
We address the problem of efficient large-scale image search with untrusted cloud servers.
Different from existing work focusing on the search efficiency,
 we also consider the image privacy (the image itself and its feature descriptors) of the image owner and
 the query privacy (the query image content and the query result) of the querier.
To guarantee the search accuracy,
 we employ the state-of-the-art large-scale image search model in the computer vision field.
In this section, we briefly review the image search model to enhance the understanding of our system.

\subsection{Image Feature Descriptor}
In the field of computer vision, many approaches have been proposed to search similar images.
\emph{Feature descriptor} is widely adopted for image similarity measurement.
Given an image $I$, interest points are detected and
 for each interest point one \emph{feature vector} $\vx_i$ is extracted
 to describe the visual characteristics around this point.
The feature descriptor of an image consists of all feature vectors extracted from it,
 denoted as $\fx:= \{\vx_1, \dots, \vx_\alpha\}$.
To achieve robust and fast image description,
 different types of descriptors are propsed, \eg, SIFT\cite{lowe2004distinctive} and SURF\cite{bay2008speeded}.
For a specific type of descriptor,
 feature vectors are usually of the same dimension,
 \eg, the feature vector of SIFT descriptor is 128-dimension.


\subsection{Voting-based Search Model}

Many modern content-based image retrieval systems
 manage millions of images, \ie, billions of high-dimension feature vectors.
Facing the big image database,
 the state-of-the-art solutions (\eg, \cite{jegou2008hamming}, \cite{jegou2009recent} and \cite{moise2013indexing})
 usually search similar images as follows:
 fist searching similar vectors for feature vectors of the querying image in an index structure,
 and then obtains matched images using a voting-based model.

Given a query descriptor  $\fx:= \{\vx_1, \dots, \vx_\alpha\}$ (descriptor of the query image $I_x$),
 and a set of of descriptors  $\{ \fy^1, \cdots, \fy^N\}$ of images in the DB,
 where $\fy^n$ is the descriptor of the $n$-th image, the voting-based search model works as follows:
 \begin{enumerate}
 \item Let the score of each image in DB be $S^n$, and initialize all scores to 0.
 \item For each feature vector $\vx_i$ of $\fx$ and each feature vector $\vy^n_j$ in DB, the score $S^n$ is increased by
 $S^n := S^n + \delta(\vx_i, \vy^n_j)$, where $\delta(\vx_i, \vy^n_j)$ is a matching function
 measuring the similarity between feature vectors $\vx_i$ and  $\vy^n_j$ based on $k$-nearest neighbors ($k$-NNs).
 Formally, the matching function is defined as
     \begin{equation}
     \delta(\vx_i, \vy^n_j) = \left\{ \begin{array}{ll}
     1 & \textrm{if $\vy^n_j$ is a $k$-NN  of $\vx_i$}\\
     0 & otherwise
     \end{array} \right.
     \end{equation}
     For the $k$-NNs search, the dissimilarity of feature vectors are
     typically measured by Euclidean distance.
 \item By ranking the image scores, images with largest scores
    are selected as the matched images.
 \end{enumerate}


Facing hundreds of feature vectors of a query image 
 and billions of feature vectors in DB,
 accurate $k$-NNs search is too expensive for most image search applications.
Approximate search greatly improves the performance but suffers from a little loss of search accuracy.
The most common way is indexing the massive feature vectors
 by partitioning them into groups via high-dimension clustering.
Given a query feature vector,
 the system firstly finds the closest cluster representative
 and fetches the contents within this cluster into the memory;
 secondly, distances between the query vector and fetched vectors
 are computed to get the $k$-NNs.
As a result,
 many clusters can be pruned quickly to accelerate the searching process.

\subsection{MapReduce Framework}
The large-scale high-dimension vector based image search 
 can be very costly for both memory and CPU even with an index structure.
MapReduce \cite{dean2008mapreduce} is a software framework for the applications which process
vast amounts of data (multi-terabyte datasets) on large clusters (thousands of
nodes) of commodity hardware in a parallel, reliable and fault-tolerant manner.
Exploiting the power of parallel computing of MapReduce,
 the image query can be handled within an acceptable response time \cite{moise2013indexing}.

%% file: he.tex

To empower the cloud to maintain the image index as well as search images in a privacy-preserving way, 
 we employ an efficient multi-level homomorphic encryption (HE) protocol presented by Xiao \textit{et al.} \cite{xiao2012efficient}. 
The protocol is defined as follows:
\begin{definition}
The multi-level homomorphic encryption is defined by three algorithms ($K,\enh,\hed$), where $K,\enh$ and $\hed$ are the key generation, encryption and decryption algorithms, and it satisfies  $\hed(\enh(m,k))=m$ given $k\leftarrow K(1^\lambda)$.
\end{definition}

We carefully review the design and security proof of this protocol,
 and make sure it is correct.
Considering it is not the contribution of this work,
 we omitted the detail explanation due to space limitation.
Notably, the homomorphic encryption has the following good properties we want to utilize in our work:

\noindent \textbf{Additive and Multiplicative Homomorphism}:
Their encryption has the following homomorphism:
\begin{displaymath}
\begin{split}
\enh(m_1,k)\cdot \enh(m_2,k)&=\enh(m_1m_2,k)\\
\enh(m_1,k)+\enh(m_2,k)&=\enh(m_1+m_2,k)
\end{split}
\end{displaymath}
\noindent This also implies the homomorphism over any polynomial function $f$, \textit{i.e.,}
\begin{displaymath}
\begin{split}
&f\left(\enh(m_1,k),\enh(m_2,k),\cdots, \enh(m_l,k)\right)\\
&=\enh\left(f(m_1,m_2,\cdots, m_l),k\right)
\end{split}
\end{displaymath}

\noindent \textbf{Key Conversion}: If we have $k=\prod_{i}k_i$ for the key $k$, the encryption has the following property:
\begin{displaymath}
\left(\prod_i k_i^{-1}\right) \cdot E(m,1) \cdot \left(\prod_i k_i\right)=E(m,\prod_i k_i)=\enh(m,k)
\end{displaymath}

\noindent This implies that one does not need to decrypt and re-encrypt the message to alter the key of a ciphertext, which is very useful in our system design. Note that a randomizer is omitted for the sake of simplicity, so it is not possible to attack this encryption via brute-force search if the ciphertext size is large enough.

\subsection{Distance Calculation via HE}\label{section:distance_on_HE}

Based on HE, we can conduct the distance calculation for two feature vectors $\vx,\vy$ on the ciphertexts as follows, where $\vx(j)$ refers to the $j$-th dimension of the feature vector $\vx$, and $\sdis(\vx,\vy)=\dis^2(\vx,\vy)$.

\begin{displaymath}
\enh\left(\sdis \left(\vx,\vy\right),k\right)=\sum_k\left(\enh\left(\vx(j),k\right)-\enh\left(\vy(j),k\right)\right)^2
\end{displaymath}

Then, the distance calculation can be outsourced to anyone who does not know the key by giving him all the ciphertexts. Since the calculation is conducted on the ciphertexts, the outsourcing does not reveal information about feature vectors or the calculation output. Hereafter, we use the notation $\phi_\sdis\left(\cdot\right)$ to denote the function which conducts distance calculation given homomorphic ciphertexts of two vectors. That is:
\begin{displaymath}
\phi_\sdis\left(\enh\left(\vx,k\right),\enh\left(\vy,k\right) \right)=\enh\left(\sdis(\vx,\vy),k\right).
\end{displaymath}


\subsection{Fixed Point Representation}\label{sec:fpr}
The numeric type of feature vectors may be real number, but the homomorphic encryption used in this paper is based on large integers, therefore we need to use integers to represent real numbers first. In PIC, we use the fixed point representation (\cite{yates2009fixed}) to represent real numbers due to its simplicity when applying elementary arithmetic operations to it.

Given a real number $a$ and a fixed precision $p$, an $m+1$-dimension binary array $A$ satisfies the following in the fixed point representation with Two's Complement:

{ \scriptsize
\begin{displaymath}
\begin{cases}
A[0]=0\wedge \sum_{k=1}^mA[k]\cdot 2^{-k+p}\approx a  & a\geq 0 \\
 A[0]=1 \wedge -\sum_{k=1}^{m-1}\left(A\left[k\right]\oplus 1\right)\cdot 2^{-k+p}-A[m]\cdot 2^{-k+m}\approx a& a<0\\
\end{cases}
\end{displaymath}
}

The minimum unit of this representation is $2^{p-m}$ (\ie, precision), and the range of this representation is $(-2^{p-1},2^{p-1})$. Then, we use the following integer to represent the real number $a$ (with some errors less than the minimum unit):

\begin{displaymath}
\fpr(a)=\begin{cases}
 \sum_{k=1}^m A[k]\cdot 2^{m-k} & \text{if }A[0]=0 \\
 -\sum_{k=1}^m \left(A\left[k\right]\oplus 1\right)2^{m-k} & \text{if }A[0]=1 \\
\end{cases}
\end{displaymath}

\noindent which is the definition of signed integer with Two's complement.

\begin{table}[h]
\caption{Fixed Point Representation Operations}\label{table:fpr_operation}
\centering
\begin{tabular}{|r|c|}
\hline
$+,-$& $\fpr(a\pm b)=\fpr(a)\pm \fpr(b)$\\
\hline
$\times$& $\fpr(a\cdot b) = {\fpr(a)\cdot \fpr(b)} / {2^{m-1-p}}$\\
\hline
\end{tabular}
\end{table}

Note that the addition/subtraction ($x\pm y$), multiplication ($x\cdot y$) and the division ($\frac{x}{y}$) are all elementary arithmetic operations closed in integer domain (\ie, $a/b$ is the quotient of $\frac{a}{b}$). We assume $m$, $p$ are pre-defined parameters based on the range and precision requirements of the application. For simplicity, we will omit the real-integer conversion and use normal arithmetic operations on real numbers in the following presentation, but the values must be converted to the fixed point representation and fixed point representation operation (Table~\ref{table:fpr_operation}) should be applied in applications.  

%% file: model.tex
We present the overview of our system design,
 threat model and the security assumption
 before further presenting design details.

\subsection{Architecture \& Entities}

\begin{figure*}
\centering
\includegraphics[width=0.8\linewidth]{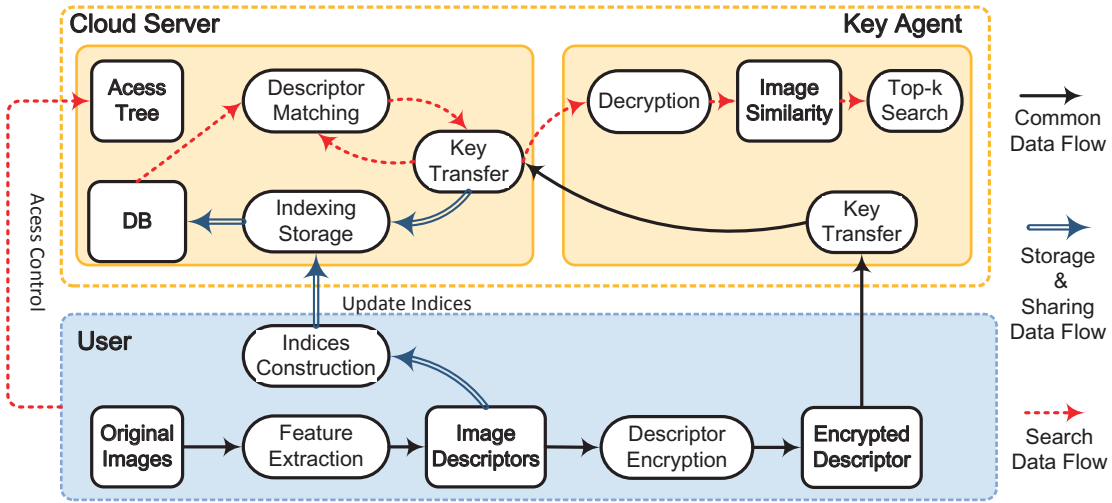}
\caption{PIC Architecture}
\label{fig:architecture}
\end{figure*}

Fig.~\ref{fig:architecture} describes the flow of our system. The system is designed to let users store, share and search images via external
cloud servers without conducting computationally heavy tasks. The entire outsourced computation is conducted on the ciphertexts directly, therefore users' image content privacy is preserved against cloud servers or other adversaries.
Specifically, for the image owner our system protects his images, images' feature descriptors and search results (similarities) from unauthorized parties (including KA and CS); for the image querier, our system also protects the querying content from unauthorized parties.
More precisely, we have the following entities in our system:
\begin{itemize}
\item \textbf{Users}: a user can be an image owner and an image querier simultaneously. An owner stores and shares his images with others by outsourcing them to the cloud servers, and a querier searches an image on the DB located at the cloud server side.

\item \textbf{Cloud Server (CS)}: The cloud server is in charge of majority of the computation and the storage throughout the system. Whenever a transaction request arrives, he processes the request by computing on the ciphertexts. CS could be any commercial cloud service providers who are willing to improve their services to attract more privacy-sensitive users.

\item \textbf{Key Agent (\KA)}: The key agent manages various secret keys such that no one within the system (including himself) learn the final key used in the encryption. Majority of the transaction between users and CS is relayed by \KA.
    KA could be any agent who is unlikely to collude with the CS or a specific user.
    For example, for crowd-sourced criminal investigation, KA can be a government-controlled agent;
    for medical image analysis, KA can be an authoritative medical organization;
    for personal image management, KA can be a reputable information security service provider.
\end{itemize}

\subsection{Threat Model}
The trusted party (TP) is introduced only to generate
the keys, who could be an auditor or a notary public and is assumed to be fully trusted.
However,
the CS and the KA who
conduct most of the computation may be motivated to
infer useful information from the outsourced computation,
and they are assumed to be semi-honest, i.e., they will
follow the protocol specification in general, but will also
try their best to harvest the content of the encrypted
communication.
In general, TP,CS and KA are well protected, so we
do not consider compromise attack in this work. Also,
although CS and KA are assumed to be semi-honest, the
probability that both of them collude with a specific user
is extremely small. Therefore, we assume that it is not
possible to have a user who colludes with both CS and
KA.

\subsection{Security Assumption}
Because we employ the homomorphic encryption in Xiao
et al. \cite{xiao2012efficient} whose security rely on the assumption that prime
factorization is hard, we also assume the following prime
factorization is hard:

\begin{definition}(Prime Factorization)
A prime factorization problem is to solve all $p_i$'s given a product of prime numbers $n = \prod_i p_i$.
\end{definition}

%% file: approach.tex
With the background and the system model, we are ready to present the detailed design of our system PIC. 
Our system supports the following four operations to serve the users: initialization, key generation \& policy announcement, image upload and privacy preserving image search with access control.

\subsection{Initialization}

Firstly, a TP picks the system parameter for the homomorphic encryption (\cite{xiao2012efficient}) and publishes it. Then, TP generates a master key $k$ to be used in the homomorphic encryption, and he finds two random keys $k_{CS},k_{KA}$ such that $k_{CS}k_{KA}=k$ and sends $k_{CS}$ and $k_{KA}$ to CS and \KA respectively via secure channel.

\subsection{Key Generation \& Policy Announcement}
Whenever a new user $u$ joins the system, TP generates three random keys $k_u,k_u',k_u''$ such that $k=k_uk_u'k_u''$. Then, he gives $k_u'$ to KA, 
$k_u''$ to CS and $k_u$ to the user $u$ via secure channel.

Then, the user defines the access policy which controls who can/cannot search on his images. The policy is described by an access tree as in CP-ABE \cite{bethencourt2007ciphertext}, and it is uploaded to CS for further access control.

Submitting raw attributes to CS will reveal the user's identity information (\cite{jung2013privacy}).
Therefore, the attributes as well as access policy should be masked before uploading.
Note that the access control policy is used as a black-box building block in our system,
 and here we present a simple policy which works with  CP-ABE as a baseline method.
When joining the system, every user describes his access policy with an access tree, but the attributes at the leaf nodes are replaced with the hashed values of the attributes.
Whenever a querier wishes to search on a group of specified users or the entire DB, the querier submits his hashed attributes to CS. CS then matches these hashed attributes with the access policies in the DB to find out the group of users that the querier is valid to search on.
We evaluate the practicality of the access policy by investigating a real social networking system (in Section\ref{sec:micro}),
 and our analysis shows that for most cases the simple access policy is sufficient and practical.
For some special cases,
 there can be other better options for the access control (e.g., anonymous IBE with predicate encryption \cite{katz2008predicate}) which achieves better anonymity, but this is not our main contribution, and we leave it as one of our future work to study.

\subsection{Image Upload}
Whenever a new user $u$ uploads some images, he first extracts the feature descriptors from them, and encrypts the descriptors using his key $k_u$ as follows:
\begin{displaymath}
\enh\left(\left\lbrace \fx_{i,1},\fx_{i,2},\cdots \right\rbrace,k_u \right)
\end{displaymath}
Then, the user needs to either create or update the index cluster for his feature descriptors by two phases. 
(1) Indices construction: cluster representatives are selected from feature vectors as centroids of clusters.
(2) Clustering: other vectors are assigned to clusters.
Various clustering techniques can be applied, 
and different clustering techniques have different performance and accuracy. 
One can simply use k-mean clustering to create the index of vectors. 
At each time the user wants to update the index, he will either re-construct the it or just incrementally append new representatives or nodes into the current clusters.

After the clusters are prepared, he appends references to the nodes in the cluster which points to the corresponding ciphertexts of feature vectors. Then, the raw index clusters are sent to CS. To reduce the communication overhead, the user can send only the change of the index cluster instead. Once he completes the update (or creation), the ciphertexts are sent to \KA.

\KA, upon receiving the ciphertexts, conducts the following operation for every ciphertext to get the ciphertexts with altered key $k_uk_u'$:
\begin{displaymath}
k_u'^{-1}\enh\left(\left\lbrace \fx_{i,1},\cdots \right\rbrace,k_u \right)k_u'= \enh\left(\left\lbrace \fx_{i,1},\cdots \right\rbrace,k_uk_u' \right)
\end{displaymath}
Then, \KA sends the altered ciphertexts to CS.

CS conducts the following operation to get the final ciphertexts:
\begin{displaymath}
\begin{split}
k_u''^{-1}\enh\left(\left\lbrace \fx_{i,1},\cdots \right\rbrace,k_uk_u' \right)k_u''
=&\enh\left(\left\lbrace \fx_{i,1},\cdots \right\rbrace,k_uk_u'k_u'' \right)\\
=&\enh\left(\left\lbrace \fx_{i,1},\cdots \right\rbrace,k \right)
\end{split}
\end{displaymath}
Then, CS merges the user $u$'s index cluster with the global one for his DB, but leaving a label to mark the owner of the cluster.

\subsection{Privacy-preserving Search with Access Control}

One image search has two phases: level-1 search and level-2 search. In the level-1 search, \KA first finds out the cluster representative in the index cluster which is closest to the querying feature vector. Then,  in the level-2 search, he finds out the $k$-nearest neighbors ($k$-NNs) of the querying feature vector within the cluster.

\noindent \textbf{Level-1 Search}

When a querier $q$ wants to search an image, he first extracts the feature descriptor (\textit{i.e.,} a set of feature vectors) from the querying image. Then, he encrypts the feature descriptor $\fx_q$  of the querying image with his key $k_q$ as $\enh\left(\fx_q,k_q\right)$, and submits the ciphertexts to \KA. Then, \KA alters the ciphertext to $\enh\left(\fx_q,k_qk_q'\right)$ and sends them to CS. CS finally alters the ciphertexts to
\begin{displaymath}\begin{split}
\enh\left(\fx_q,k_qk_q'k_q''k_{CS}^{-1}\right)&=\enh\left(\fx_q,kk_{CS}^{-1}\right)\\
&=\enh\left(\fx_q,k_{KA}\right)
\end{split}\end{displaymath}

Besides, the querier also uploads his hashed attributes to CS. CS then searches all users' access policies and finds out the users that the querier is valid to search on their images. Then, CS computes the following altered ciphertexts, where $\{\vy_o\}$ refers to the set of representative feature vectors (cluster leaders) in previously found owners' index clusters, and all $\{\vy_o\}$ are stored in the form of ciphertexts encrypted by $k$:

\begin{displaymath}
\begin{split}
k_{CS}\enh\left(\left\lbrace\vy_o\right\rbrace,k \right)k_{CS}^{-1}
=&\enh\left(\left\lbrace\vy_o\right\rbrace, kk_{CS}^{-1}\right)\\
=&\enh\left(\left\lbrace\vy_o\right\rbrace, k_{KA}\right)
\end{split}
\end{displaymath}
After all the altered ciphertexts are ready, CS computes the $\phi_\sdis(\cdot)$ function (Section \ref{section:distance_on_HE}) for every pair of\\ $\left(\enh\left(\vx,k_{KA}\right),\enh\left(\vy,k_{KA}\right)\right)$, where $\vx\in \fx_q$ is a feature vector belonging to the descriptor of the querying image, and $\vy\in\{\vy_o\}$. Then, CS achieves the pairwise encrypted distances, which are sent to \KA.

\KA is able to decrypt the distances since the ciphertexts are encrypted under his key $k_{KA}$. After decrypting the distances, he finds out the nearest neighbor for every $\vx\in\fx_q$.

\noindent \textbf{Level-2 Search}

After finding the nearest neighbor among the representatives for every $\vx\in\fx_q$,
KA further requests the distances between the $\vx$ and all the vectors within the nearest neighbor's cluster.

Upon receiving the request, CS generates the following altered ciphertexts using the key conversion (Section \ref{section:HE}) and $\phi_\sdis(\cdot)$ function as aforementioned, where $\left\lbrace \vy_c \right\rbrace$ is the set of vectors within the nearest neighbor's cluster:
\begin{displaymath}
\left\lbrace\left\lbrace\enh\left(\sdis\left(\vx,\vy\right),k_{KA}\right)\right\rbrace_{\forall\vy\in\{\vy_c\}}\right\rbrace_{\forall \vx\in\fx_q}
\end{displaymath}
Then, he sends these ciphertexts of distances as well as the image IDs associated with the feature vectors in the ciphertexs to \KA. \KA decrypts the ciphertexts and determines the $k$-NNs among $\left\lbrace\enh\left(\sdis\left(\vx,\vy\right),k_{KA}\right)\right\rbrace_{\forall\vy\in\{\vy_c\}}$ for each $\vx\in\fx_q$. Based on the distances and the corresponding image IDs, he calculates the score $S^n$ of all images (Section \ref{sec:prelim}) appearing in the image IDs sent from CS and returns the image ID with the highest score to the querier. The querier then retrieves the encrypted image from the DB. One can further apply oblivious transfer (will be described in Section \ref{sec:review}) to prevent the CS from inferring the query result by monitoring its memory access.

In this level-2 search, if CS does not find $k$ ciphertexts within one cluster, he also chooses the next nearest neighbor among the representatives and sends the corresponding ciphertexts to \KA as well. This is repeated until he finds out at least $k$ ciphertexts to return to \KA.

%% file: system.tex
The basic system we proposed in the previous section works fine in some cases, 
 \eg people recognition in a face image collection,
 but its performance is degraded in more general cases as we will discuss below. 
Therefore, we present several improvements for our scheme to boost the performance, 
 and finally formally describe the advanced scheme with these optimizations in this section.

\subsection{Dealing with Complicated Images}
A feature descriptor usually contains a set of high-dimension feature vectors, \eg 128 dimensions for SIFT.
Since each dimension is a 64-bit real value, when we consider the encryption,
each feature vector's size becomes 64KB because each dimension is encrypted with a $4\times 4$ matrix with 256-bit integers.
Then, the number of feature vectors in an image determines the size of its feature descriptor.
For the face recognition, 9 feature vectors are enough to conduct an accurate search
 because face models are well developed.
However,
for complicated image with hundreds of feature vectors,
the size of ciphertexts is not acceptable for many mobile devices.
Therefore, we further optimize our system using more compact image descriptor to reduce the communication overhead.

We adopt frequency vector of visual words as image descriptor(\cite{jegou2010improving,sivic2003video}).
It quantizes the space of feature vectors to obtain a visual dictionary of size $n$.
An image can be represented with the frequency histogram of \textit{visual words},
 by choosing the nearest word for each of its feature vectors.
Then the feature descriptor is significantly reduced to 
 one $n$-dimension frequency vector.
The similarity between images can be computed by the scalar product of two frequency vectors.

To minimize the accuracy loss caused by quantizing the feature vector space to a discrete one,
 we weight the frequency vector with term frequency inverse document frequency (tf-idf, \cite{sivic2003video}).
Based on tf-idf, given a dictionary of $n$ words and an image dataset of $N$ images,
the weighted frequency vector of an image $I$ is $\vf_I=\left(w_1,w_2,\cdots,w_m\right)$, and each weight $w_i$ is:
$w_i=\frac{f_{i,I}}{n_I}\log \frac{N}{f_i}$,
where $f_{i,I}$ is the frequency of the lexicographic (in the visual dictionary) $i$-th feature vector in the image $I$, $n_I$ is the total number of feature vectors in $I$, $f_i$ is the frequency of the $i$-th feature vector among $N$ images.
The experimental results in peer work(\cite{sivic2003video}) and our evaluation result (Section~\ref{sec:accuracy}) show that using this weighted visual word instead of exact feature vectors brings a negligible accuracy loss.
Therefore, we use this weighted frequency vector with visual dictionary as our feature descriptor in our advanced scheme.

\CUTTH{
Overall,
 the the system design requires to explore the problem that:
 to design privacy-preserving search schemes adapted to the Map-Reduce paradigm;
  and find the best setting to maximize search quality for the given response time
 and the privacy protection level requirements.
}

\CUTTH{
\subsection{System Parameters}

The key parameters governing the computation and storage performance of this approach include:
\begin{enumerate}
\item $L$: the recursive level $L$ leads to the expected number of distance computation for a query being $(L+1)N^{\frac{1}{L+1}}$ and larger $L$ increases costs in random I/O requests to external memory;
\item $C$: the number of clusters created, which determines the average cluster size; larger $C$ results in a longer clustering process but a smaller search delay (the cluster size should be considered with the IO parameters);
\item $\alpha$: in the tree of representatives,
 each representative is associated to its closest $\alpha$ parent representatives;
\item  $\beta$: $\beta$ governs how many leaves are exhaustively searched for each query.
\end{enumerate}

There are some analysis about the parameter setting for
 best search quality \cite{chierichetti2007finding, gudmundsson2010large},
however,
 the privacy protection blocks change both the computation cost
 and dist cost of common systems,
 which makes the result of exiting studies cannot be applied directly.
Moreover,
 we should consider the effect of parameters on privacy protection level.
 }

\CUTTH{
\subsection{Mapper and Reducer}

Phase 1 is cheap and can be completed by the client in a plane way.
Phase 2 is computation costly and requires a parallelization and distribution.
If Step 2 is done by the cloud,
 then it requires privacy-preserving distance calculation and comparison.

The steps for privacy-preserving preprocessing with Map-Reduce:
\begin{enumerate}
\item The tree of representative is constructed by the client.
\item the tree and the rest vectors are encrypted and uploaded to the cloud.
\item the cloud clusters the rest vectors in a privacy preserving way
 in a Map-Reduce paradigm:
 the encrypted tree and data blocks of vectors
 are sent to various nodes; each node assigns its subset of vectors
 to clusters.
\end{enumerate}

The Map and Reduce tasks:
each mapper loads the tree and read-assign-emit every descriptor in its subset.
The emitted key of each descriptor is the identifier of its assigned cluster.
the reduce tasks receive results grouped by cluster identifier from the shuffle
 and propagate the data to disk to form the clusters.

Giving a set of query descriptors of an query image
 there are  phases to search its matching images:
\begin{enumerate}[Phase 1]
\item $k$-nn search: find the $k$ nearest neighbors search for each query descriptors.
\item voting: each nearest neighbor votes for the image it was extracted from, and the image
with the most votes are returned as an match to the query.
\end{enumerate}
The main building block is privacy-preserving Euclidean Distance calculation, voting
 and comparison.

The steps for privacy-preserving query processing with Map-Reduce:
\begin{enumerate}
\item The descriptors are extracted from the query image.
\item The query descriptors are encrypted and uploaded to the cloud.
\item The cloud searches the $\beta$ closes clusters for each query descriptors.
\item The cloud searches $k$ nearest neighbors for $m$ query descriptors in a Map-Reduce diagram.
\item The cloud uses the $k$-nn search results to vote and find the match image.
\end{enumerate}

The Map and Reduce tasks:
each mapper receives the whole $m$ query descriptors ordered by their closest cluster identifiers
and start start to process its data blocks (the clustered descriptors).
Its blocks are loaded and their contents are processed only if clusters in their contents
 are needed by at least one query descriptor.
Each mapper maintains the $k$-nn tables for all query descriptors
 concerned with the current cluster under analysis
 and a series of records are emitted.
The reduce tasks receives $k$-nn tables and merges them for the final $k$-nn table for all query descriptors.
}

\subsection{Leveraging the Parallel Computation}


We further expedite the search process by applying the MapReduce framework. To do so, the search must support parallelism. Looking at the level-1 search and the level-2 search, one can easily find both searches (finding the NN among the representatives and the k-NNs within the cluster) can be done by arbitrary number of mappers and one reducer. The ciphertexts of the feature vectors in DB and the querying vectors will be given to the mappers at CS, who conducts the homomorphic operations and the key modification to generate the ciphertexts of the distances. Then, they are sent to the reducer at \KA who decrypts and sorts the distances.

However, there is a problem if our approach is directly implemented with a MapReduce framework. 
Unless the data is stored with a special format (\eg, bucketized), 
sorting limits the number of reducers to one because the mappers do not emit the (key,value)=(image ID, distance) pairs in a sorted order. 
However, since the sorting order depends on the querying feature vector, the number of reducers will be always 1. 
This is a great bottleneck since the reducer needs to wait for all (key,value) pairs emitted from all mappers before it continues to sort. 
To solve this bottleneck, we can return all the neighbors within a certain threshold distance instead of the exact NN in the level-1 search or the k-NNs in the level-2 search. By doing so, we can theoretically have arbitrary number of reducers to finish the search task in a parallel manner. 
If the threshold is chosen such that not enough results are found, one can easily use a binary search manner to adjust the threshold until enough results are found for further processing.

\subsection{Choosing a Good Clustering Method}
In fact, the clustering in the image upload operation heavily affects the performance and the accuracy,
 especially when it works with the MapReduce framework. 
\emph{Extended cluster pruning approach with recursion} \cite{gudmundsson2010large, chierichetti2007finding}
 provides a promising way for efficient clusters construction and queries processing,
 and can be fitted appropriately in the MapReduce framework.
Here we briefly review this approach and apply it in the advanced scheme.

Given a set $\vV=\{\vv_1, \vv_2, \dots, \vv_N\}$ of $N$ vectors in $d$-dimension space
 and a positive integer $L$.
The cluster pruning based approach searches
 $k$-nearest neighbors for a query vector $\vq$ via two phases:
 \emph{preprocessing} and \emph{query processing}.
During the preprocessing,
 $C$ vectors are chosen as \emph{representatives} at random and denoted by $\vr_1, \dots, \vr_C$.
$C$ is determined by $C=N/S$, where $S$ is the target size of each cluster.
The $C$ representatives are organized in a multi-level hierarchy of $L$ levels.
Each other vector in $\vV$ traverses the tree and
 is attached to its closest representatives at the bottom of the tree.
The vectors are partitioned into $C$ clusters $\vC_1, \dots, \vC_C$
 with a logarithmic complexity.
The clusters will be stored on a hard disk while the tree of representatives
 is small enough to be fitted in a memory, which can be processed efficiently by a mapper.
Hereafter, we denote the representative of each cluster as $\vl_i$, which is also called as the \emph{leader} of this cluster.

During the query processing,
 by navigating down the tree of cluster leaders,
 the nearest leader of a query vector $\vq$
 is searched.
Then the corresponding bottom cluster is fetched
  and the $k$-NNs for $\vq$ are sought inside this cluster.
Obviously, the $k$-NNs for $\vq$ within the cluster is not always the global $k$-NNs of $\vq$ among all feature vectors. 
Then, to improve the search accuracy, one may leverage the state-of-the-art soft-assignment and multi-probe techniques.
The soft-assignment allows each vector to be assigned to its $\alpha$ nearest representatives. 
The multi-probe allows each search to probe $\beta$ closest clusters simultaneously to search the overall $k$-NNs among these clusters.

\subsection{Advanced Scheme}

Only the image upload and the privacy-preserving image search with access control will be changed in the advanced scheme. 
The changed operations are as follows:

\subsubsection{Image Upload}
The user $u$ firstly extracts the feature vectors from every image that he wants to upload to CS. 
Then, he uses $k$-mean clustering to find the $k$ clusters among all his vectors, and sets all the centroids as elements in the visual dictionary of size $k$. Hereafter, we use $\fd_u$ to denote user $u$'s dictionary.
After $\fd_u$ is generated, user $u$ also calculates the $k$-dimension weighted frequency vector of each image $I$ as $\vf_I$, and uses CP-ABE to encrypt all parameters in the weight function ($\{f_{i,I},f_i\}_i,N$) as well as $\fd_u$: 
\begin{displaymath}
\abe\left(\fd_u,\{f_{i,I},f_i\}_i=1,\cdots,k,N\right).
\end{displaymath}
User $u$ sends this ciphertext to CS, and encrypts all weighted frequency vectors as:
$\enh\left(\left\lbrace\vf_I\right\rbrace_I,k_i\right)$.
Then, he uses aforementioned extended cluster pruning approach with recursion to construct an index tree of all $\vf_I$'s, where each node contains the reference to the corresponding encrypted vector. The raw index tree is sent to CS (if only updated, only the changed parts are sent to CS), and the encrypted ciphertexts are sent to \KA.

\KA, upon receiving the ciphertexts, conducts the key modification to alter the ciphertexts to:
\begin{displaymath}
\enh\left(\left\lbrace\vf_I\right\rbrace_I,k_uk_u'\right).
\end{displaymath}
Then, \KA sends the new ciphertexts to CS, who conducts another key modification to achieve the final ciphertexts:
\begin{displaymath}
\enh\left(\left\lbrace\vf_I\right\rbrace_I,k_uk_u'k_u''\right)=\enh\left(\left\lbrace\vf_I\right\rbrace_I,k\right).
\end{displaymath}

\subsubsection{Privacy-preserving Search with Access Control}
Same as the basic scheme, we also have two phases in a search.

\vspace{0.2in}
\noindent \textbf{Level-1 Search}

When a querier $q$ wants to search an image, he first extracts the feature vectors from the querying image. Then, he retrieves the $\abe\left(\fd_o,\{f_{i,I},f_i\}_i=1,\cdots,k,N\right)$'s of all image owner $o$'s that he wants to search on, and decrypts the dictionaries as well as the parameters with his attributes. Then, he looks up each dictionary to create a frequency vector of his querying image for each $\fd_o$. Then, he calculates $m$ weighted frequency vectors with the decrypted parameters, which are encrypted as:
$\enh\left(\{\vf\}_{\text{all owners}},k_q\right)$.
This is relayed by \KA (after altering the key to $k_qk_q'$) and finally arrives CS who finally alters the key to $k_qk_q'k_q''=k$. Then, CS finds out the index trees of the users that querier $q$ wishes to search on, and conducts the following key modification, where $\{\vy_o\}$ refers to the set of frequency vectors referenced (via ciphertext) by the cluster leaders in those index trees:
\begin{displaymath}
\enh\left(\left\lbrace\vy_o\right\rbrace, kk_{CS}^{-1}\right)=\enh\left(\left\lbrace\vy_o\right\rbrace, k_{KA}\right).
\end{displaymath}
Then, CS computes the $\phi_D(\cdot)$ function for every pair of $\left(\enh\left(\vx,k_{KA}\right),\enh\left(\vy,k_{KA}\right)\right)$ where $\vx$ is the frequency vector of the querying image, and $\vy\in \{\vy_o\}$. The outputs of the function are the pairwise encrypted distances, which are sent to \KA.

\KA proceeds with MapReduce framework. He sends the ciphertexts to his mappers, each mapper decrypts the distances and emits the (key,value)=(image ID,distance) to his reducers, and the reducers find out the distances above a threshold $\theta$, which corresponds to the distances of $\vx$'s nearest neighbors.

\noindent \textbf{Level-2 Search}

After finding the NNs, \KA further requests the distances between $\vx$ and all vectors under those NNs in the index trees.

Upon receiving the request, CS generates the encrypted distances using the key modification as well as the $\phi_D(\cdot)$ function. 
They are transmitted to \KA who sends those ciphertexts to his mappers to let them decrypt the distances. 
The (image ID,distance) pairs are sent to his reducers, and they finally find out the images whose distances are above another threshold $\theta'$.

%% file: ana.tex
The security of a system is given by the security of its weakest link.
Since the participating adversaries are more powerful than non-participating adversaries and colluding adversaries
are more powerful than a single adversary,
 we analyze the security of our system in the worst case scenario: colluding participating adversaries, including CS, \KA and users (TP is fully trusted).
 Note that CS and \KA do not collude with each other, and a user can collude with at most one party of CS and \KA. 
Therefore, we have the following two cases: colluding user and CS; colluding user and \KA.

It is already formally proved in \cite{xiao2012efficient} that the homomorphic encryption is secure against colluding user \& CS or colluding user \& \KA, hence no party in the system can infer the plaintext from the ciphertext without knowing the key. 
We further analyze the extra information leakage in our system.

\noindent \textbf{Colluding user and \KA}

Colluding user and \KA only learn $k_u,k_u'$ during the key generation, index construction or the index update. 
During the image search, \KA receives a number of encrypted distances $\enh\left(\sdis(\vx,\vy),k_{KA}\right)$ 
which are encrypted with $k_{KA}$. Then, \KA can decrypt the distances, and the colluding user (a querier in this case) will know all the distances between the encrypted feature vectors in the database and his querying feature vectors. However, no ciphertexts of vectors are sent to \KA or user.
Therefore, the user only learns a number of distances, which is not useful to infer the images in CS's database.
All the user gets is only the query result, which is the ciphertext of the matched image. 
Therefore, colluding user and CS do not gain useful information except the valid search result.

\noindent \textbf{Colluding user and CS.}

Colluding user and CS only learn $k_u,k_u''$ during the key generation. During the index forest construction and the index update, CS may share the ciphertext\\ $\enh\left(\{\fx_{i,1},\fx_{i,2},\cdots\},k_uk_u'\right)$ received from \KA with user,
from which the user can use the key modification to achieve the ciphertext
$\enh\left(\{\fx_{i,1},\fx_{i,2},\cdots\},k_u'\right)$.
However, since $k_u'$ is unknown to both parties, it is not possible to infer $\fx_{i,1},\fx_{i,2},\cdots$ from the ciphertexts. During the image search, CS may also share the ciphertexts of the distance with the user, but neither CS nor the user is able to infer the distance from the ciphertext since none of them know the key $k_{KA}$.

%% file: eva.tex
In this section,
 we first present the implementation of PIC and the datasets used in the experiments,
 then we evaluate the performance of our basic scheme using SIFT feature vectors
 and advanced scheme using weighted frequency vector comprehensively.
For simplicity,
 in the following statement,
 we denote the basic scheme as PIC-sfv and the advanced scheme as PIC-wfv.

\subsection{System Implementation}
Our implementation of PIC includes both both cloud side and client side.
On the cloud side,
 each of \KA and CS consists of a cluster of computers with distributed file system (Hadoop HDFS)
 and MapReduce architecture (Hadoop MapReduce).
In the experiments,
 we use four PCs with Intel Core i3-3240 CPU (3.4GHz) and 4G RAM for each cluster.
Each cluster has one name node and three data nodes,
 which is a small but full-featured data center to demonstrate our design,
 and the performance can be greatly improved when using a large data center \cite{sangroya2012mrbs, sangroya2012benchmarking}.
On the client side,
 we implement our system for both Windows OS laptops and Android phones.
In the experiments,
 we use a laptop (ThinkPad X1) with Intel Core i7-2620M CPU (2.7GHz) and 4GB RAM,
 and a mobile phone (HTC G17) with 1228Hz CPU, 1GB RAM.
There is also a trusted party (TP) in charge of key generation.
TP is a single PC with the same hardware as the node of the cluster.

We implement attribute based access control with the PBC (Pairing-Based Cryptography) library \cite{pbc}.
We develop all other cryptographic components are implemented by Java,
 including the fix point arithmetic and the multi-level homomorphic encryption \cite{xiao2012efficient}.
We use three commonly used descriptors to evaluate the practicality of our system,
 including 128-dimension SIFT descriptor \cite{lowe2004distinctive},
 64-SURF and 128-SURF \cite{bay2008speeded}.
Nevertheless, our schemes are also compatible with other image descriptors.
The descriptor extraction
 is implemented using the OpenCV library for Window and Android.
Due to the space limitation,
 in this paper, we only present the results of the most popular descriptor, 128-SIFT,
 and 64-SURF and 128-SURF achieve similar performance as that using 128-SIFT.

\subsection{Image Collections and Queries}
To explore the performance of our approach in real-life image applications,
 we evaluate our system with two popular image datasets.
\begin{itemize}
\item  \textbf{INRIA Holiday dataset (Holiday)} \cite{jegou2008hamming}
 contains 1491 personal holidays photos in high resolution (most are 2560*1920).
 There are 6767563 SIFT feature vectors of dimensionality 128 extracted from those images.
 The dataset contains 500 image groups, each of which represents a distinct scene or object.
 For the search experiments, the query is a photo of a scene,
 and the goal is to return other $k$ photos of this scene.
\item \textbf{Flickr image dataset (Flickr1M)} \cite{Flickr1M} contains one million diverse images from Flickr
 with 1.4 billion pre-computed SIFT feature vectors in total.
 For the search experiments, the query is randomly selected images,
 and $k$ most similar images are returned.
 \end{itemize}

\begin{figure}[t!]
\centering
\includegraphics[width=0.8\linewidth, clip]{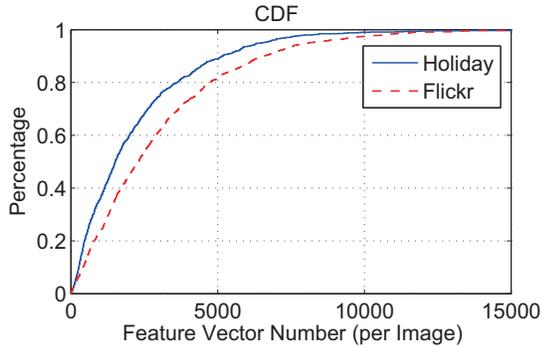}
\caption{CDF of feature vector number of images from two datasets.}\label{fig:vector-num}
\end{figure}

We analyze the feature vector number of each image in two datasets.
As shown in Fig.~\ref{fig:vector-num},
The mean feature vector number is 2,200 and 2,988 for Holiday and Flickr1M respectively.
Half images have less than 2000 feature vectors
 and $80\%$ images have less than 5000 feature vectors.

We evaluate PIC-sfv using SIFT feature vectors
 from both datasets.
For the evaluation of PIC-wfv,
 1000 visual words are learned from 6K randomly selected images of Flickr1M as the vocabulary.
And a 1000-dimension weighted frequency vector is generated for each image by clients.

\subsection{Parameter Selection}
Before we present our evaluation results,
 we first discuss the selection of our experiment parameters.

\subsubsection{Search Parameters Selection}
There are three parameters governing the computation, communication and search performance for both schemes,
 including: (1) The number $C$ of created clusters, which determines the delay, overhead and accuracy of a search.
 Larger $C$ results in a longer clustering procedure and lower search accuracy, but a smaller search delay.
 There are some analysis on the parameter selection for
 optimal search quality \cite{chierichetti2007finding, gudmundsson2010large},
 and we follow them to set $C=\sqrt{N}$, resulting in $2\sqrt{N}$ homomorphic distance calculations,
 where $N$ equals number of feature vectors for PIC-sfv and number of weighted frequency vectors for PIC-wfv.
 (2) The number $k$ of nearest neighbors, which influences the search accuracy.
Voting-based methods are not very sensitive to $k$ for large collections \cite{jegou2010improving},
 and we set $k=5$ in our evaluation, which produces good search results (Section~\ref{sec:accuracy}).
 (3) The work nodes number $N_{node}$ in the parallel computation.
 Existing work such as \cite{sangroya2012mrbs} and \cite{sangroya2012benchmarking} have studied
  the performance gain as the number of work nodes increases.
 Based on those work,
  it is easy to estimate the performance of our system with more computing nodes.

For PIC-wfv,
 another key parameter is the size of the visual vocabulary $v$.
Larger $v$ yields more accurate search result.
It also determines the dimension of the weighted frequency vector for each image.
As a result, the computation and communication cost increase linearly
 with $v$.
Based on the ground truth of Holiday,
 we set $v=1000$ to optimize the search accuracy ($93.4\%$)
 with acceptable overhead.


\subsubsection{Security Parameters Selection}
For the multi-level homomorphic encryption, there are two parameters $\lambda$ and $m$ governing the security level
 and overhead of the system.
The system can withstand an attack with up to $m \ln \text{ploy}(\lambda)$ chosen plaintexts,
 while the communication cost increases linearly with $m\lambda$ and computation cost increases exponentially with $m\lambda$.
As a result,
 there is a tradeoff between security and efficiency.
In our experiments,
 we choose $m$=2 and $\lambda=128$,
 which allows $m\ln\lambda^{10}\sim 97$ plaintext attacks with good computation and communication cost.
This might be dangerous for local applications,
 because normally adversaries are allowed to access decryption oracle for polynomial times,
 but our system will remain safe since the cloud server will not allow such 'decryption oracle access' for several times.

\begin{figure*}[t!]
\begin{minipage}[b]{0.45\linewidth}
\centering
\includegraphics[width=0.75\textwidth, clip]{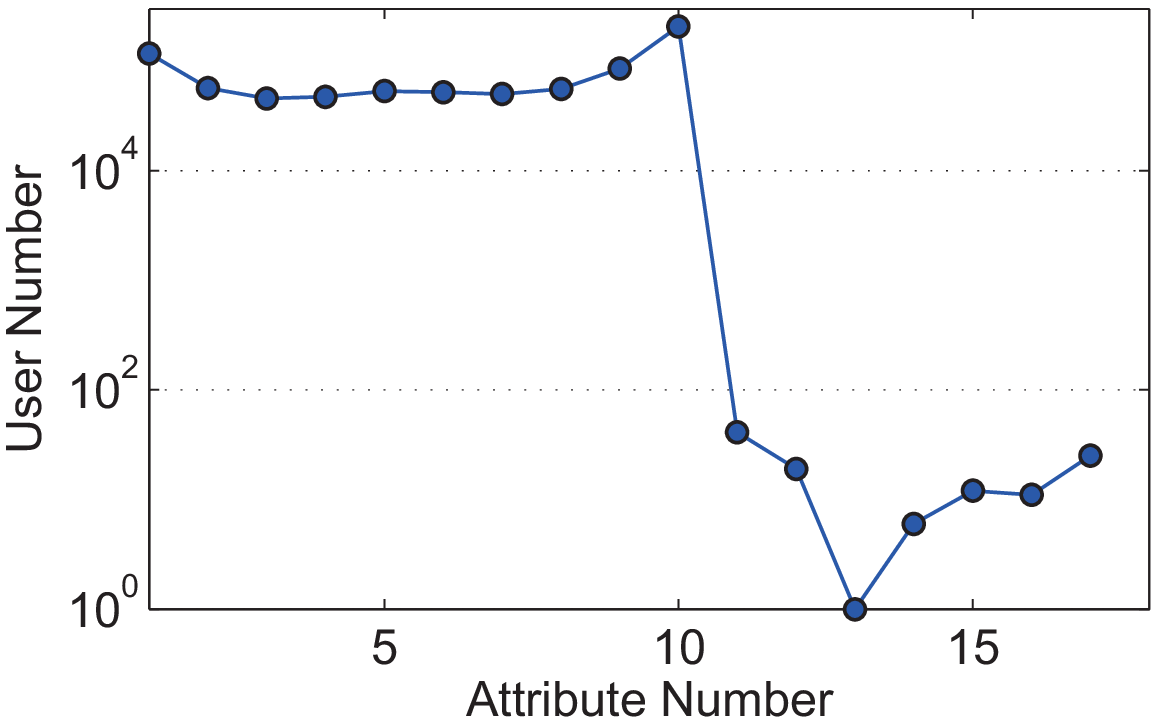}
 \caption{Users' attribute number distribution of Tencent Weibo.}
\label{fig_tag_num}
\end{minipage}
\hfill
\begin{minipage}[b]{0.45\linewidth}
\centering
\includegraphics[width=0.75 \textwidth, clip]{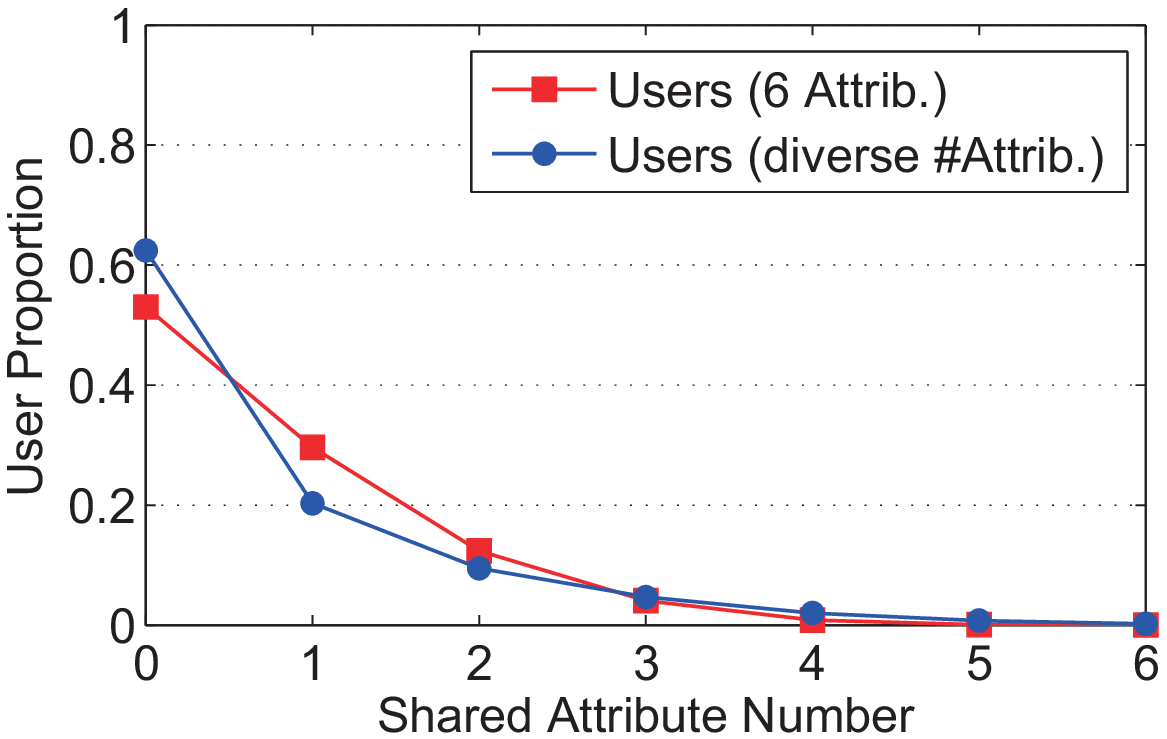}
\caption{The proportion of Tencent Weibo users having common attributes, except user ID which is unique for every user.
 The blue line is for all users with different number of attributes. The red line is for those users
 who have 6 attributes. }
\label{fig_profile_collision}
\end{minipage}
\end{figure*}

\subsection{Micro-Analysis for Each Operation}
\label{sec:micro}

We analyze the additional computation and communication overhead introduced by our system except the image process related overhead in this subsection.
Due to the space limitation,
 we omit the evaluation results on the Android phone,
 whose runtime is the same order of magnitude as that on the laptop.

\begin{table}[t]
\caption{Computation cost for each operation using a single PC. (s)
 $\lambda=128$, $m=2$,the SIFT-vec is 128-dimension SIFT feature vector, the Freq-Vec is 1000-dimension weighted frequency feature vector.}
\label{table:time}
\centering
{\scriptsize
\begin{tabular}{|l|c|c|c|}
\hline
\multicolumn{4}{|c|}{Common Operations}\\
\hline
& Mean & Min & Max\\
\hline
Feature Extract & $8.7$& $1.56$ & $12.03$\\
\hline
Parameter Init & 0.002 & 0.002 & 0.003\\
\hline
Master Key Gen& <0.001  & <0.001 & <0.001\\
\hline
KA\&CS Keys Gen&  0.001 & 0.001 & 0.001\\
\hline
Triple Keys Gen&  0.002 & 0.002 & 0.003\\
\hline
Access Tree Gen& <0.001  & <0.001 & <0.001\\
\hline
Decrypt Distance& <0.001  & <0.001 & <0.001\\
\hline
\multicolumn{4}{|c|}{Operations of PIC-sfv}\\
\hline
Encrypt One SIFT-Vec&  0.034 & 0.03  & 0.047\\
\hline
Key Modification (per SIFT-Vec)&  0.052 & 0.048& 0.063\\
\hline
Homomorphic Euclidean Distance&  0.031 & 0.029& 0.033\\
\hline
\multicolumn{4}{|c|}{Operation of PIC-wfv}\\
\hline
& Mean & Min & Max\\
\hline
Freq-Vec Gen& 0.70 & 0.005  & 3.84 \\
\hline
Encrypt One Freq-Vec & 1.93 & 1.59 & 2.67\\
\hline
Key Modification (per Freq-Vec)& 0.29  & 0.25 & 0.85\\
\hline
Homomorphic Dot Product & 0.20 & 0.18 & 0.33 \\
\hline
\end{tabular}}
\end{table}

\textbf{(1) Computation Overhead}

The runtime of each operation is summarized in Table~\ref{table:time}
 and then the detailed analysis is presented as follows.

\noindent \textbf{Initialization}: In this step, it requires a selection of system parameters for the homomorphic encryption 
and a selection of three random homomorphic encryption keys $k,k_{CS},k_{KA}$, which are $4\times 4$ matrices, such that $k=k_{CS}k_{KA}$.
Both operations are executed at the TP side 
and our results show that the system initiation takes less than 5ms in total, which is negligible.

\noindent \textbf{Key Generation \& Policy Announcement}:
In this step, it requires TP to select three random keys $k_i,k_i',k_i''$ such that $k=k_ik_i'k_i''$, which costs less than 3ms.
Besides, it also involves an owner's access tree generation.
We evaluate the performance of our access control methods based on the profile data of Tencent Weibo \cite{tencent},
 which is one of the largest social networking platform in China.
This dataset has $2.32$ million users' personal profiles,
 including their year of birth, gender, graduate school, profession and other tags.
There are $770166$ different attributes in total.
Each user has $6$ attributes in average and $20$ attributes at most, as shown in Fig.~\ref{fig_tag_num}.
Fig.~\ref{fig_profile_collision} illustrates the proportion
 of users sharing different number of common attributes.
60\% users have no common attribute with others,
 about 20\% users share one common attribute with others, and 98\% users share less than four common attributes
 with others.
Our analysis suggests that a small access tree with limited attributes (\eg 6 attributes), is enough to
 narrow down the size of authorized users (only $0.2\%$ is valid),
 and its generation time is only $8\times 10^{-3}$ms.
Given the access tree,
 the runtime to authenticate the attributes of a querier is less than 1ms.
So the cost for access control is negligible.



\begin{figure*}[ht!]
\begin{minipage}[b]{0.45\linewidth}
\centering
\includegraphics[width=0.8\textwidth, clip]{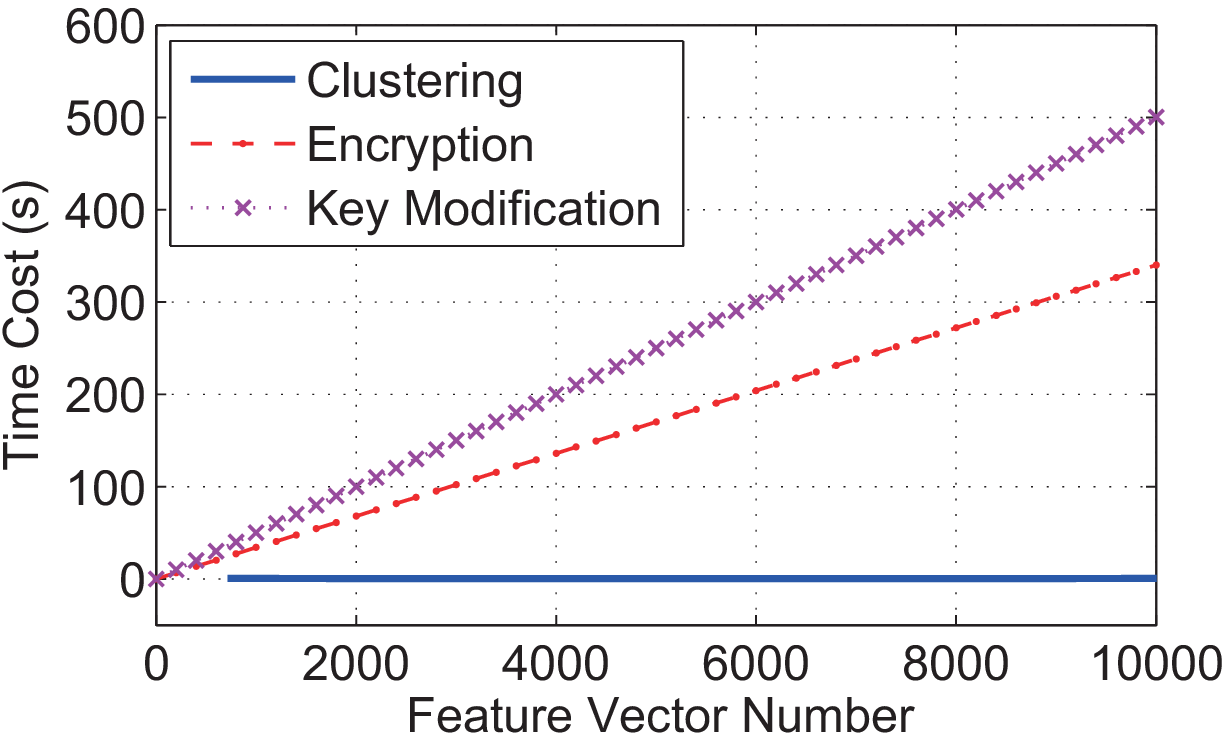}
 \caption{Runtime of upload operation VS. SIFT feature vector Number (PIC-sfv).}
\label{fig_upload_vec}
\end{minipage}
\hfill
\begin{minipage}[b]{0.45\linewidth}
\centering
\includegraphics[width=0.8\linewidth, clip]{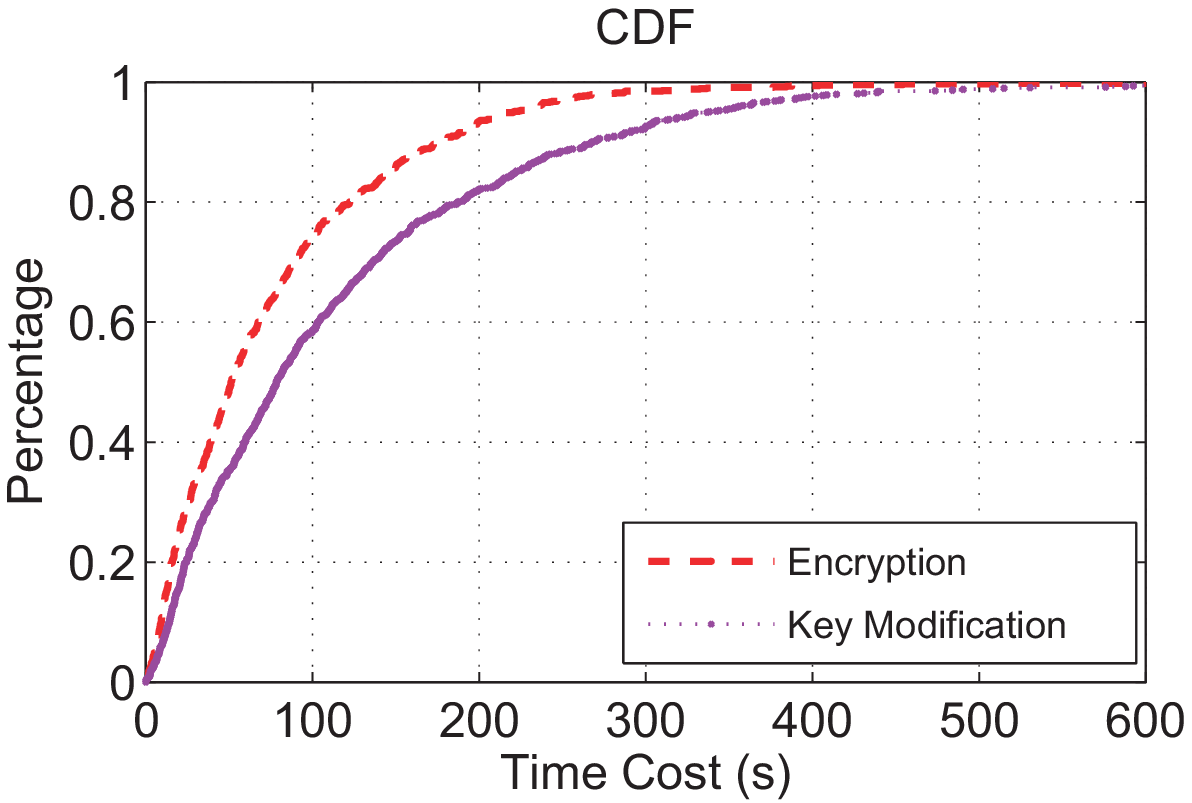}
\caption{CDF of runtime of upload operation for each image in two image sets (PIC-sfv).}\label{fig:upload-time-basic}
\end{minipage}
\end{figure*}

\noindent \textbf{Image Upload}:
For PIC-sfv,
 the cluster construction is first executed by each owner,
 and the runtime increases linearly with the feature vector number (Fig.~\ref{fig_upload_vec}).
For two image sets,
 it takes about 20 seconds to cluster feature vectors of 100 randomly selected images.
Then the owner encrypts every descriptor using his key $k_i$.
As shown in Table~\ref{table:time}, it takes 34ms to encrypt each 128-dimension feature vector.
The runtime to encrypt the descriptor of each image depends on
 the its feature vector number as depicted in Fig.~\ref{fig_upload_vec}.
Fig.~\ref{fig:upload-time-basic} shows the runtime distribution to encrypt one image from two images sets,
 which is 75s in average.
After the ciphertexts are sent to \KA, \KA conducts a key modification to alter the encryption key of the ciphertexts.
As shown in Fig.~\ref{fig:upload-time-basic},
 it takes 155s to modify the key of an image in average.
The key modification at CS side is the same operation as \KA's one and thus omitted.

\begin{figure*}[ht]
\begin{minipage}[b]{0.45\linewidth}
\centering
\includegraphics[width=0.8\textwidth, clip]{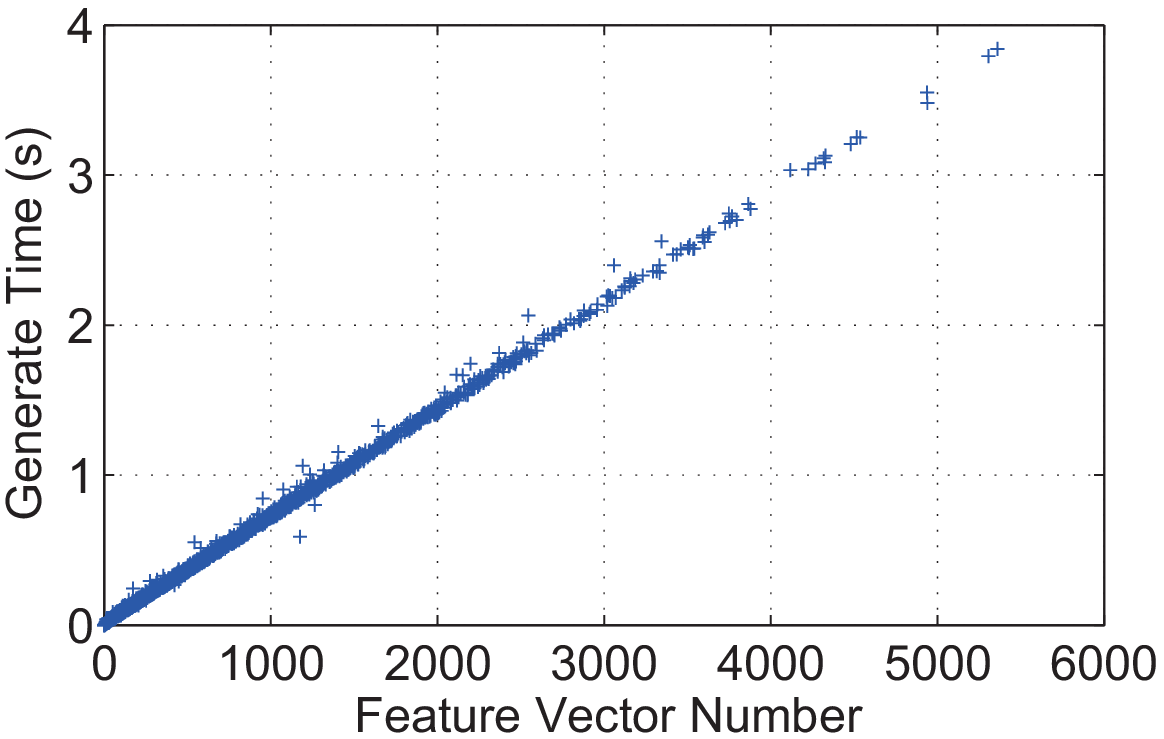}
 \caption{1000-dimension weighted frequency vector generation for one image VS. feature vector number of this image (PIC-wfv).}\label{fig:wfv-gen}
\end{minipage}
\hfill
\begin{minipage}[b]{0.45\linewidth}
\centering
\includegraphics[width=0.8\linewidth, clip]{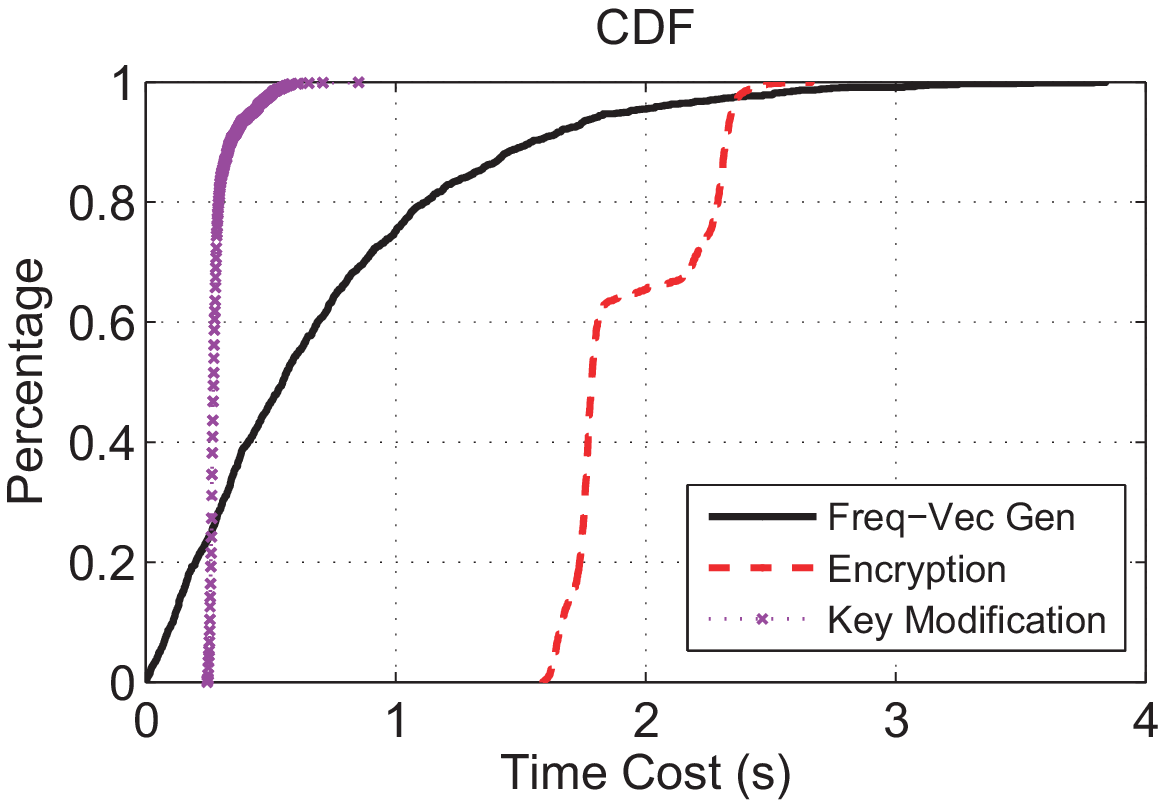}
\caption{CDF of runtime of upload operation for each image in two image sets (PIC-wfv).}\label{fig:upload-time-advanced}
\end{minipage}
\end{figure*}

For PIC-wfv,
 based on the visual word vocabulary,
 the owner generates weight frequency vector for each image,
 whose runtime is proportional to the feature vector number of this image (Fig.~\ref{fig:wfv-gen}).
For two image sets,
 it takes 0.7s per image in average.
Then the owner encrypts the weighted frequency vector of each image,
 whose runtime is depicted in Fig~\ref{fig:upload-time-advanced}.
It takes less than 1.5s to encrypt one image.
Fig.~\ref{fig:upload-time-advanced} also presents the time cost
 of key modification for each image by \KA (or CS),
 which is about 0.29s.
The evaluation shows that the advance scheme based on the weight frequency vector
significantly improve the computation efficiency.

\noindent \textbf{Privacy-preserving Image Search with Access Control}:
The querier first encrypts the image descriptor (SIFT feature vectors or one weighted frequency vector),
 whose run time is the same as the one in the image uploading (Fig.~\ref{fig:upload-time-basic} and Fig.~\ref{fig:upload-time-advanced}).
The key modifications by \KA and CS are also the same.
Besides, in PIC-sfv, CS needs to compute the ciphertexts of all squared Euclidean distances.
Using a single machine, each distance takes 31ms and the whole time cost depends on the number of query feature vectors
 and the size of the searched vector collection in DB, \ie $(\#cluster + cluster\ size)\times \#feature vector\times 31$ms.
In PIC-wfv, CS computes the ciphertexts of all dot products,
 each takes 200ms by one machine and the whole time cost just depends on the image collection size, \ie $(\#cluster + cluster\ size) \times 200$ms.
After CS prepares all the ciphertexts, \KA decrypts them to achieve the distances or dot products,
 and sort them to find out the NN.
For both schemes, each decryption requires less than 1ms and it takes about 0.5s to decrypt 1000 distances.
We use quick-sort to sort the distances, but we omit the run time analysis since this is a standard sorting method.
For Holiday,
 without MapReduce functions,
 using PIC-sfv each query (with more than 3000 feature vectors) averagely takes about 100 machine-hours to find the matching image among the 1491 images;
 using PIC-wfv each query takes about 15 machine-seconds.
Deploying our  MapReduce implementations on the 4-node small data center,
 the delay is reduced to about 17 hours for PIC-sfv
 and reduced to 4s for PIC-wfv.

As a brief summary, both PIC-sfv and PIC-wfv have same initialization delay
 and similar runtime for index construction and frequency vector generation.
However, PIC-wfv uses only one 1000-dimension weighted frequency vector to represent each image,
 while in PIC-sfv the descriptor of each image is a set of 128-dimension SIFT feature vectors.
When the size of feature vectors is small,
 two scheme has comparable performance.
But as the feature vector size increases,
 PIC-sfv's overhead increases linearly while PIC-wfv keeps the runtime almost a constant.
Moreover, for the privacy-preserving image search,
  our schemes work excellently with MapReduce framework to reduce the response time.

\textbf{(2) Communication Overhead}

\begin{table}[h]
\caption{Communication cost of each data structure. (KB)
 $\lambda=128$, $m=2$,the SIFT-vec is 128-dimension SIFT feature vector, the Freq-Vec is 1000-dimension weighted frequency feature vector.}
\label{table:size}
\centering
{\small
\begin{tabular}{|l|c|c|c|}
\hline
\multicolumn{4}{|c|}{Common Data}\\
\hline
& Mean & Min & Max\\
\hline
Key &  0.84 & 0.19 & 1.4\\
\hline
Access Tree & 0.18 & 0.03 & 0.63\\
\hline
Encrypted Distance& 0.56 & 0.41 & 0.65\\
\hline
\multicolumn{4}{|c|}{Data of PIC-sfv}\\
\hline
Encrypted SIFT-Vec&  64 & 62.8  & 65.1\\
\hline
\multicolumn{4}{|c|}{Data of PIC-wfv}\\
\hline
& Mean & Min & Max\\
\hline
Encrypted Freq-Vec& 580 & 578.9  & 581.2 \\
\hline
\end{tabular}}
\end{table}

We first summarize the size of the transmitted data structure in Table~\ref{table:size}.
Then we analyze the communication cost for each operation.

\noindent \textbf{Initialization}:
TP needs to send key $k_{CS}$ and $k_{KA}$ to CS and \KA respectively, and the mean size of a single key is 0.84KB.

\noindent \textbf{Key Generation \& Policy Announcement}:
TP sends $k_i,k_i'k_i''$ to the user $i$, CS and \KA respectively, and each key's size is 0.84KB.
Based on the user attributes of Tencent Weibo,
 the size of the access tree with 6 attributes is about 0.2KB.

\noindent \textbf{Image Upload}:
The user informs CS of the change in the index cluster, but this is almost negligible.
Main communication overhead comes from the ciphertexts transmission.
For PIC-sfv,
 uploading the encrypted feature vectors incurs $\#feature\ vector \times 64$KB data transmission.
For PIC-wfv,
 the ciphertexts (encrypted weighted frequency vector) size of each image is 580KB,
 which is constant.
Similarly, \KA also needs to send out ciphertexts of the same size to CS.

\noindent \textbf{Privacy Preserving Image Search with Access Control}:
First, the querier encrypts the query descriptor and sends the corresponding ciphertexts to \KA,
which is \\$\#feature\ vector \times 64$KB for PIC-sfv and 580KB for PIC-wfv.
\KA also sends the same amount of ciphertexts to CS.
For PIC-sfv,  in the level-1 search,
after CS computes the encrypted distances for all representatives, the encrypted distances are sent back to \KA,
 the size is 0.56KB each and the whole size is $\#cluster \times 0.56$KB.
During the level-2 search, CS computes the encrypted distances for all vectors within the NN's cluster,
 and sends them to \KA ($cluster\ size \times 0.56$KB).
For PIC-wfv,
 similarly, CS computes and send all encrypted dot products to \KA, whose size is $(\#cluster + cluster\ size ) \times 0.56$KB.
For Holiday,
 the transmitted data between CS and \KA during search is about 2800KB for two rounds search of PIC-sfv
 and 43KB for PIC-wfv.

\subsection{Macro-Analysis for Each Entity}
We analyze the overall computation overhead for each entity in this subsection. Similarly, we only analyze the additional overhead except that for the image processing.


\noindent \textbf{Cloud Server}: The computational delay at CS side during the image uploading comes from key modification.
In PIC-sfv, it is $\#feature\ vector \times 52$ms (Fig.~\ref{fig_upload_vec})
 and about 200s for $80 \%$ images (Fig.~\ref{fig:upload-time-basic}).
Using PIC-wfv, it takes about 1.2s for $80\%$ images.
During the image search,
 the computation cost of CS comes from the homomorphic distance calculation.
For a single node, the cost is $(\#cluster + cluster\ size)\times \#feature\ vector\times 31$ms
 using PIC-sfv and $(\#cluster + cluster\ size)\times 200$ms using PIC-wfv.

\noindent \textbf{Key Agent}: \KA experiences the same delay as CS during the image upload.
The computation cost of \KA during search comes from decrypting all distances and ranking them to find the NN.
The run time is 0.5s to process 1000 distances.

\noindent \textbf{Client}: The computational delays occur during the system join, image upload and the image search for an ordinary user.
The system join cost is negligible (about 3ms).
In PIC-sfv,
 the clustering cost is negligible compared to the encryption cost, and encryption cost is $\#feature\ vector \times 34$ms (Fig.~\ref{fig_upload_vec}).
So the computational delay of upload is about 100s for $80\%$ images from the two image sets (Fig.~\ref{fig:upload-time-basic}).
Similarly, for PIC-wfv, the computational delay during upload for  $80\%$ images is only about 2.2s
and 1.5s in average(Fig.~\ref{fig:upload-time-advanced}).
During search,
 the client does nothing but waits for the search result from the cloud
 and the delay is the summation of computational delay at CS and \KA.

\begin{figure}[ht]
\centering
\includegraphics[width=0.8\linewidth, clip]{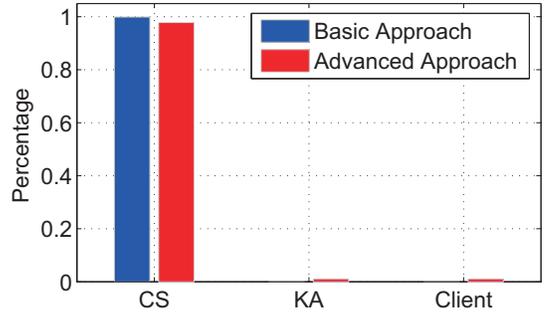}
\caption{Computation overhead distribution among CS, KA and client for a single query .}\label{fig:overhead}
\end{figure}

In summary,
 for a querier, after providing a query image, the response time is mainly the computational delay accumulation of image encryption,
 two key modifications,  homomorphic distance calculation, distance decryption and ranking.
When the querier can access all 1491 images of Holiday on the cloud,
 the average response time of each query is about 17 hours and 8 minutes for PIC-sfv
 and 7.21 seconds for PIC-wfv.
Here the delay can be greatly reduced
 as the cloud scales up from 4 nodes to hundreds of nodes.
Besides,
 compared to exiting multi-part secure computation based methods,
 by our approach,
 during the search no interaction is required for the client
 and \textbf{97\%} computation is carried out by CS, leaving KA and clients very limited overhead (Fig.~\ref{fig:overhead}).


\CUTXY{
The transmission for system join is omitted in the following analysis,
 since it is a one-time operation for each entity.

\noindent \textbf{Client}: The data transmission of the client is mainly
  uploading the ciphertexts.
For each image,the size is $\#feature\ vector \times 64$KB for PIC-sfv,
  and 580KB for the advance approach.

\noindent \textbf{Key Agent and Cloud Server}:
During upload, \KA transmits the ciphertexts uploaded by the client to CS.
In the search of PIC-sfv,
 CS sends two  $(\#cluster + cluster\ size) \times 0.56$KB packages of encrypted distances to \KA.
In the search of PIC-wfv,
 CS sends $\#image \times 0.56$KB distance data to \KA.
}

\subsection{Performance Comparison}
\label{sec:accuracy}

\paragraph{Compare with alternative methods.}
\cite{xiao2012efficient} has compared its cost against the well-known homomorphic encryption scheme of Gentry \cite{Gentry2009fully}.
  Gentry's scheme needs more than 900 seconds to add two 32 bit numbers, and  more than 67000 seconds for the multiplication,
  but the cost for \cite{xiao2012efficient} is only 0.1 ms and 108 ms respectively.
  The reason is that Gentry's fully homomorphic encryption is based on the ''learning with errors'' (LWE) problems in lattice system,
   which allows users to apply as many multiplications as they want on the ciphertexts.
   \cite{xiao2012efficient} is based on number theory and group theory, which only supports a limited number of homomorphic operations,
   but is sufficient for our application.
  So, \cite{xiao2012efficient} is much more practical and compact.
We also realize private Euclidean distance computation using a partial homomorphic encryption (Paillier encryption)
 in the SMC manner (e.g. the method used in \cite{katz2008predicate}).
Using the same computer and test images, the Paillier-based method (128-bit) takes about 0.5s for feature vector encryption
 and 0.18s for homomorphic distance computation. But in our work, they take only 0.034s and 0.031s respectively.
The comparison shows the computation efficiency of our system.

\paragraph{Search Accuracy.}
First,
 we evaluate the search accuracy of our approaches according to the ground truth of
 500 queries of Holiday.
With $k$=5 (five nearest neighbors are fetched),
 the accuracy of PIC-sfv is $95.2\%$
 and of PIC-wfv is $93.4\%$.
Here the accuracy is the success rate of 500 queries.
A result is success if the returned $k$ images contain
 at least one image from the same scene as the query image.
So, our solution achieves privacy-preserving without sacrificing the search accuracy.
When a vocabulary is learnt,
 PIC-wfv provides the similarly good accuracy as PIC-sfv.

\paragraph{Linear Search vs. Our Approaches with MapReduce.}
\begin{figure}[h]
\centering
\includegraphics[width=0.8 \linewidth, clip]{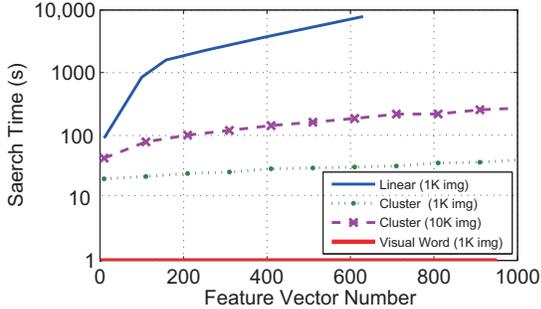}
\caption{Search time using different approaches VS. feature vector number of the query image.
Two different size searched image collections (1K and 10K) are selected randomly from Flickr1M.}\label{fig:search-time-vec}
\end{figure}

We implement the SIFT feature vector based scheme and the visual word based scheme using the
 Hadoop MapReduce framework to accelerate the search.
To study the search efficiency of different schemes,
 we compare the computational overhead of our two schemes using a 4-node cluster
 with the one of conducting a linear search using a single computer on the raw images data.
The comparison is presented at Fig.~\ref{fig:search-time-vec},
 which shows the overhead is reduced an order of magnitude by the SIFT feature vector based approach.
Furthermore, the visual word based approach keeps the overhead a second-level constant.

Then we evaluate the improvement caused by adapting our privacy-preserving search to MapReduce framework.
We run the 500 queries on the encrypted Holiday data set.
With a single computer,
 the cluster based approach takes about 100 hours for each query,
 and the visual word based approach takes 15 seconds for each query.
When running on the 4-node small cluster with our MapReduce adaptive implementation,
 the runtime reduces from 100 to 17 hours  and from 15 seconds to 4s respectively.
With a large cluster, the improvement will be much bigger.

\paragraph{With vs. Without Privacy Protection.}
By running the visual word based search on Holiday on the 4-node cluster in two cases,
 where all computation in the first case is conducted on the raw image data
 and all computation in the second case is conducted on ciphertexts as our system design.
In the first case, the average response time for each query is 1.8s
 and in the second case, that is 4s.
The result shows that our system provides good protection to the image privacy
 with little extra overhead.

As a conclusion,
 leveraging the power of indexing and MapReduce,
 our system design improves the search performance greatly
 while keeping a high search accuracy and well protected privacy.

%% file: related.tex
\subsection{Image Indexing and Search}
A lot of work address the problem searching for similar images,
 and most of them are based on the local invariant descriptors.
Typically, high-dimension descriptors are extracted from the interest regions of images
 to represent their visual characteristics.
Different types of descriptors are proposed
 to achieve efficient and accurate image matching,
 \eg, SIFT \cite{lowe2004distinctive}
 and SURF\cite{bay2008speeded}.
The 128-dimension SIFT is the most widely used descriptor for image search
 due to its distinctiveness and computational efficiency.
An accurate approach to search similar images
 is to measure the distance between images' descriptors
 and conduct nearest neighbor search.
But facing billions of high-dimension vectors,
the accurate nearest neighbor search is too expensive.
Various feature vector indexing approaches are designed to
 boost the search efficiency.
A common way is using clustering
 to prune many clusters during the $k$-NN search, \eg, kd-tree \cite{muja2009fast} and cluster pruning \cite{chierichetti2007finding}.
Among various recent research work, extended cluster pruning \cite{gudmundsson2010large}
 provides a promising way for efficient cluster construction and query processing,
 which outperforms the comparable solutions like p-sphere tree \cite{goldstein2000contrast} and rank aggregation \cite{dwork2001rank}.
Besides clustering,
 Sivic and Zisserman \cite{sivic2003video} introduce
 the bag-of-features (BOF) for image search,
 which quantizes feature vectors into limited visual words
 using the k-means algorithm.
An image is then represented by a visual word frequency vector.
Based on visual words,
 a lot of work speed up the search process \eg, \cite{jegou2010improving}.
As commercial data center get more and more popular,
 it is also possible to improve large-scale image search using
 computer clusters.
MapReduce\cite{dean2008mapreduce} provides an efficient programming framework
 for processing large data set in parallel.
There are some work using MapReduce to accelerate the indexing and
 search process, \eg, \cite{moise2013indexing}.
However, few of the image indexing and search systems consider the privacy protection of the image owner and the querier.


%

\subsection{Image Privacy Protection}
To protect privacy information in the image,
 there are some applications simply encrypting the image or blacking out private contents, \eg, human faces.
\cite{newton2005preserving} removes facial characteristics from the video frame
 to protect the face privacy of individuals in video surveillance.
P3 \cite{ra2013p3} proposes a privacy preserving photo sharing scheme
 by separating an image into private part and public part, and encrypts the private part directly.
Many access control mechanisms are proposed to enable data owners to determine
 who can access their outsourced private data.
For example, the data encrypted by ABE \cite{sahai2005fuzzy, goyal2006attribute, guo2012paas} can only be decrypted by the user whose
 attributes satisfy the access rule.
\cite{bethencourt2007ciphertext} presents a Ciphertext-Policy Attribute-Based Encryption (CP-ABE)
 which keeps encrypted data confidential even if the storage server is untrusted.
And the CP-ABE is improved in \cite{waters2011ciphertext}.
Those work provide privacy protection to image storage,
 leaving the processed image of limited use
 and no image search is supported for the private image.
GigaSight \cite{simoens2013scalable} proposes an Internet system for collection of
 crowd-sourced video from mobile devices,
 which blacks out sensitive information from video frames.
For search purpose,
 each frame is analyzed by computer vision code to obtain tags as the index.
However, it cannot support image similarity based search and the indices also expose sensitive information.

\subsection{Privacy Preserving Search}
Searching on encrypted (SoE) data was introduced by Song \textit{et al.} \cite{song2000practical}.
It allows users to store their encrypted data on untrusted server and later search
 the data by keywords in a privacy-preserving manner, \ie, both the keywords and data will not be revealed.
 Reza \textit{et al.} present a thorough discussion on the framework of searchable symmetric encryption \cite{reza2004sse}.
Chang \textit{et al.} improve the security and efficiency of SoE \cite{chang2005privacy}.
Golle \textit{et al.} develop a method supporting conjunctive keyword search \cite{golle2004secure}.
Lu \textit{et al.} \cite{lu2011privacy} propose a scheme allowing the data owner to delegate users to conducting content-level fine-grained
private search and decryption.
Cong, et.al. \cite{cong2012keyword} extend the framework to ranked keyword search.
Recently, many work are dedicated to improve the search efficiency and flexibility , \eg, \cite{wang2012enabling,li2012toward, wang2012achieving, renprivacy, sun2014protecting}.
Chen et al. \cite{chen2012large} design a system supporting large-Scale privacy-preserving mapping of human genomic sequences on
hybrid clouds using a well designed hash-based mechanism.
But those work mainly target text data and focus on keywords search by examining the occurrences of the searched terms (or words).
They are not suitable for content-based image search since they cannot measure the distance between encrypted high-dimension feature vectors.
Moreover, they usually reveal the search results to the cloud.

\subsection{Secure Multi-part Computation and Homomorphic Encryption}
Privacy-preserving similarity measurement as well as search can be achieved using secure multi-party computation (SMC) \cite{yao1982protocols}.
SMC enables multiple parties to jointly compute a function over their inputs,
 while at the same time keeping these inputs private.
There are some work addressing private distance computing among two parties
 using SMC methods \cite{lindell2004proof}.
Homomorphic encryption, \eg, Pallier and Elgamal,
 allows user to conduct the computation on
 the ciphertexts and obtain the ciphertext of the result,
 which matches the result of computation on the plaintexts.
There are some work providing privacy-preserving image matching
 using classic homomorphic encryptions.
For example, \cite{erkin2009privacy} and \cite{sadeghi2010efficient}
 allow a client privately search for a specific face image
 in the face image database of a server.
Those methods provide privacy protection to the query image
 as well as the outcome of the matching algorithm,
 but the result is not secure against the service provider.
All those privacy-preserving search mechanisms based on SMC
 and homomorphic encryption require rounds of online interactions with data owners during the search.
Besides, computation cost of those methods are very expensive and none of them can be
scaled to address large-scale image sets.

Recently, Xiao \textit{et al.} propose an efficient homomorphic encryption
 protocol for multi-user system. \cite{xiao2012efficient}
It is a non-circuit based symmetric-key homomorphic encryption scheme,
 whose security is equivalent to the large integer factorization problem.
We employ this protocol to design our system.

%

%% file: conclusion.tex
We have presented a system PIC, which enables privacy-preserving content-based search
 on large-scale outsourced images.
With our design, the image owner can determine who is valid to search and access his images.
In the image searching protocol, majority of the computationally intensive image matching jobs are outsourced to the cloud side,
but the image privacy is preserved.
The content of images at cloud's DB, the query result and even the query itself are kept secret to anyone else but the image owner or the querier.
To further expedite the search process, we introduced the index structure and made the entire search process paralleled.
We implemented our prototype system using the Hadoop MapReduce framework in a cluster of 4 computers,
 and our experiment results show the efficiency and applicability of our system.

%% file: PMR.bbl
\begin{thebibliography}{10}
\providecommand{\url}[1]{#1}
\csname url@samestyle\endcsname
\providecommand{\newblock}{\relax}
\providecommand{\bibinfo}[2]{#2}
\providecommand{\BIBentrySTDinterwordspacing}{\spaceskip=0pt\relax}
\providecommand{\BIBentryALTinterwordstretchfactor}{4}
\providecommand{\BIBentryALTinterwordspacing}{\spaceskip=\fontdimen2\font plus
\BIBentryALTinterwordstretchfactor\fontdimen3\font minus
  \fontdimen4\font\relax}
\providecommand{\BIBforeignlanguage}[2]{{%
\expandafter\ifx\csname l@#1\endcsname\relax
\typeout{** WARNING: IEEEtran.bst: No hyphenation pattern has been}%
\typeout{** loaded for the language `#1'. Using the pattern for}%
\typeout{** the default language instead.}%
\else
\language=\csname l@#1\endcsname
\fi
#2}}
\providecommand{\BIBdecl}{\relax}
\BIBdecl

\bibitem{Boston}
``Boston marathon investigation,''
  http://www.wired.com/2013/04/boston-crowdsourced.

\bibitem{korn1996fast}
F.~Korn, N.~Sidiropoulos, and C.~Faloutsos, ``Fast nearest neighbor search in
  medical image databases,'' 1996.

\bibitem{tagare1997medical}
H.~D. Tagare, C.~C. Jaffe, and J.~Duncan, ``Medical image databases a
  content-based retrieval approach,'' \emph{Journal of the American Medical
  Informatics Association}, vol.~4, no.~3, pp. 184--198, 1997.

\bibitem{jegou2007contextual}
H.~Jegou, H.~Harzallah, and C.~Schmid, ``A contextual dissimilarity measure for
  accurate and efficient image search,'' in \emph{CVPR}.\hskip 1em plus 0.5em
  minus 0.4em\relax IEEE, 2007.

\bibitem{jegou2010improving}
H.~J{\'e}gou, M.~Douze, and C.~Schmid, ``Improving bag-of-features for large
  scale image search,'' \emph{IJCV}, vol.~87, no.~3, pp. 316--336, 2010.

\bibitem{ion2011home}
I.~Ion, N.~Sachdeva, P.~Kumaraguru, and S.~{\v{C}}apkun, ``Home is safer than
  the cloud!: privacy concerns for consumer cloud storage,'' in
  \emph{Proceedings of the Seventh Symposium on Usable Privacy and
  Security}.\hskip 1em plus 0.5em minus 0.4em\relax ACM, 2011, p.~13.

\bibitem{weinzaepfel2011reconstructing}
P.~Weinzaepfel, H.~J{\'e}gou, and P.~P{\'e}rez, ``Reconstructing an image from
  its local descriptors,'' in \emph{CVPR}.\hskip 1em plus 0.5em minus
  0.4em\relax IEEE, 2011.

\bibitem{lowe2004distinctive}
D.~G. Lowe, ``Distinctive image features from scale-invariant keypoints,''
  \emph{IJCV}, vol.~60, no.~2, 2004.

\bibitem{jegou2009recent}
H.~Jegou, M.~Douze, and C.~Schmid, ``Recent advances in large scale image
  search,'' in \emph{Emerging Trends in Visual Computing}.\hskip 1em plus 0.5em
  minus 0.4em\relax Springer, 2009.

\bibitem{gudmundsson2010large}
G.~{\TH}. Gudmundsson, B.~{\TH}. J{\'o}nsson, and L.~Amsaleg, ``A large-scale
  performance study of cluster-based high-dimensional indexing,'' in
  \emph{Proceedings of the international workshop on Very-large-scale
  multimedia corpus, mining and retrieval}.\hskip 1em plus 0.5em minus
  0.4em\relax ACM, 2010, pp. 31--36.

\bibitem{daneshi2011image}
M.~Daneshi and J.~Guo, ``Image reconstruction based on local feature
  descriptors,'' 2011.

\bibitem{d2012beyond}
E.~d'Angelo, A.~Alahi, and P.~Vandergheynst, ``Beyond bits: Reconstructing
  images from local binary descriptors,'' in \emph{International Conference on
  Pattern Recognition}.\hskip 1em plus 0.5em minus 0.4em\relax IEEE, 2012.

\bibitem{Keepsafe}
``Keepsafe application,'' https://www.getkeepsafe.com/.

\bibitem{facebook_off}
D.~M. Reporter, ``Facebook to switch off controversial facial recognition
  feature following data protection concerns,''
  http://www.dailymail.co.uk/news/article-2207098/Facebook-switch-controversial-facial-recognition-feature-following-data-protection-concerns.html,
  Sep. 2012.

\bibitem{song2000practical}
D.~X. Song, D.~Wagner, and A.~Perrig, ``Practical techniques for searches on
  encrypted data,'' in \emph{Security and Privacy}.\hskip 1em plus 0.5em minus
  0.4em\relax IEEE, 2000, pp. 44--55.

\bibitem{golle2004secure}
P.~Golle, J.~Staddon, and B.~Waters, ``Secure conjunctive keyword search over
  encrypted data,'' in \emph{Applied Cryptography and Network Security}.\hskip
  1em plus 0.5em minus 0.4em\relax Springer, 2004, pp. 31--45.

\bibitem{lu2011privacy}
Y.~Lu and G.~Tsudik, ``Privacy-preserving cloud database querying,''
  \emph{Journal of Internet Services and Information Security}, vol.~1, no.~4,
  pp. 5--25, 2011.

\bibitem{cong2012keyword}
C.~Wang, N.~Cao, K.~Ren, and W.~Lou, ``Enabling secure and efficient ranked
  keyword search over outsourced cloud data,'' \emph{IEEE TPDS}, vol.~23,
  no.~8, pp. 1467 -- 1479, 2012.

\bibitem{yao1982protocols}
A.~C.-C. Yao, ``Protocols for secure computations,'' in \emph{FOCS}, vol.~82,
  1982, pp. 160--164.

\bibitem{lindell2004proof}
Y.~Lindell and B.~Pinkas, ``A proof of yao's protocol for secure two-party
  computation.'' \emph{IACR Cryptology ePrint Archive}, vol. 2004, p. 175,
  2004.

\bibitem{parno2013pinocchio}
B.~Parno, J.~Howell, C.~Gentry, and M.~Raykova, ``Pinocchio: Nearly practical
  verifiable computation,'' in \emph{S\&P}.\hskip 1em plus 0.5em minus
  0.4em\relax IEEE, 2013.

\bibitem{erkin2009privacy}
Z.~Erkin, M.~Franz, J.~Guajardo, S.~Katzenbeisser, I.~Lagendijk, and T.~Toft,
  ``Privacy-preserving face recognition,'' in \emph{Privacy Enhancing
  Technologies}.\hskip 1em plus 0.5em minus 0.4em\relax Springer Berlin
  Heidelberg, 2009, pp. 235--253.

\bibitem{sadeghi2010efficient}
A.-R. Sadeghi, T.~Schneider, and I.~Wehrenberg, ``Efficient privacy-preserving
  face recognition,'' in \emph{Information, Security and Cryptology}.\hskip 1em
  plus 0.5em minus 0.4em\relax Springer Berlin Heidelberg, 2010, pp. 229--244.

\bibitem{xiao2012efficient}
L.~Xiao, O.~Bastani, and I.-L. Yen, ``An efficient homomorphic encryption
  protocol for multi-user systems.'' \emph{IACR Cryptology ePrint Archive},
  vol. 2012, p. 193, 2012.

\bibitem{bay2008speeded}
H.~Bay, A.~Ess, T.~Tuytelaars, and L.~Van~Gool, ``Speeded-up robust features
  (surf),'' \emph{Computer vision and image understanding}, vol. 110, no.~3,
  pp. 346--359, 2008.

\bibitem{jegou2008hamming}
H.~Jegou, M.~Douze, and C.~Schmid, ``Hamming embedding and weak geometric
  consistency for large scale image search,'' in \emph{ECCV}.\hskip 1em plus
  0.5em minus 0.4em\relax Springer, 2008.

\bibitem{moise2013indexing}
D.~Moise, D.~Shestakov, G.~Gudmundsson, and L.~Amsaleg, ``Indexing and
  searching 100m images with map-reduce,'' in \emph{International conference on
  multimedia retrieval}.\hskip 1em plus 0.5em minus 0.4em\relax ACM, 2013.

\bibitem{dean2008mapreduce}
J.~Dean and S.~Ghemawat, ``Mapreduce: simplified data processing on large
  clusters,'' \emph{Communications of the ACM}, vol.~51, no.~1, pp. 107--113,
  2008.

\bibitem{yates2009fixed}
R.~Yates, ``Fixed-point arithmetic: An introduction,'' \emph{Digital Signal
  Labs}, vol.~81, no.~83, p. 198, 2009.

\bibitem{bethencourt2007ciphertext}
J.~Bethencourt, A.~Sahai, and B.~Waters, ``Ciphertext-policy attribute-based
  encryption,'' in \emph{S\&P}.\hskip 1em plus 0.5em minus 0.4em\relax IEEE,
  2007.

\bibitem{jung2013privacy}
T.~Jung, X.-Y. Li, Z.~Wan, and M.~Wan, ``Privacy preserving cloud data access
  with multi-authorities,'' in \emph{IEEE INFOCOM}, 2013.

\bibitem{katz2008predicate}
J.~Katz, A.~Sahai, and B.~Waters, ``Predicate encryption supporting
  disjunctions, polynomial equations, and inner products,'' in \emph{Advances
  in Cryptology--EUROCRYPT 2008}.\hskip 1em plus 0.5em minus 0.4em\relax
  Springer, 2008, pp. 146--162.

\bibitem{sivic2003video}
J.~Sivic and A.~Zisserman, ``Video google: A text retrieval approach to object
  matching in videos,'' in \emph{ICCV}.\hskip 1em plus 0.5em minus 0.4em\relax
  IEEE, 2003.

\bibitem{chierichetti2007finding}
F.~Chierichetti, A.~Panconesi, P.~Raghavan, M.~Sozio, A.~Tiberi, and E.~Upfal,
  ``Finding near neighbors through cluster pruning,'' in \emph{Proceedings of
  the twenty-sixth ACM SIGMOD-SIGACT-SIGART symposium on Principles of database
  systems}.\hskip 1em plus 0.5em minus 0.4em\relax ACM, 2007, pp. 103--112.

\bibitem{sangroya2012mrbs}
A.~Sangroya, D.~Serrano, and S.~Bouchenak, ``Mrbs: A comprehensive mapreduce
  benchmark suite,'' \emph{LIG, Grenoble, France, Research Report RR-LIG-024},
  2012.

\bibitem{sangroya2012benchmarking}
------, ``Benchmarking dependability of mapreduce systems,'' in \emph{Symposium
  on Reliable Distributed Systems}.\hskip 1em plus 0.5em minus 0.4em\relax
  IEEE, 2012.

\bibitem{pbc}
``Pbc library,'' http://crypto.stanford.edu/pbc/.

\bibitem{Flickr1M}
``Flickr1m dataset,'' http://www.multimedia-computing.de/wiki/Flickr1M.

\bibitem{tencent}
``Tencent weibo,'' http://t.qq.com/.

\bibitem{Gentry2009fully}
C.~Gentry, ``Fully homomorphic encryption using ideal lattices,'' in
  \emph{STOC}.\hskip 1em plus 0.5em minus 0.4em\relax ACM, 2009.

\bibitem{muja2009fast}
M.~Muja and D.~G. Lowe, ``Fast approximate nearest neighbors with automatic
  algorithm configuration.'' in \emph{VISAPP (1)}, 2009, pp. 331--340.

\bibitem{goldstein2000contrast}
J.~Goldstein and R.~Ramakrishnan, ``Contrast plots and p-sphere trees: Space
  vs. time in nearest neighbour searches,'' in \emph{VLDB}, 2000.

\bibitem{dwork2001rank}
C.~Dwork, R.~Kumar, M.~Naor, and D.~Sivakumar, ``Rank aggregation methods for
  the web,'' in \emph{International Conference on World Wide Web}.\hskip 1em
  plus 0.5em minus 0.4em\relax ACM, 2001.

\bibitem{newton2005preserving}
E.~M. Newton, L.~Sweeney, and B.~Malin, ``Preserving privacy by de-identifying
  face images,'' \emph{TKDE}, vol.~17, no.~2, pp. 232--243, 2005.

\bibitem{ra2013p3}
M.-R. Ra, R.~Govindan, and A.~Ortega, ``P3: Toward privacy-preserving photo
  sharing,'' in \emph{NSDI}.\hskip 1em plus 0.5em minus 0.4em\relax USENIX,
  2013.

\bibitem{sahai2005fuzzy}
A.~Sahai and B.~Waters, ``Fuzzy identity-based encryption,'' in \emph{Advances
  in Cryptology--EUROCRYPT 2005}.\hskip 1em plus 0.5em minus 0.4em\relax
  Springer, 2005, pp. 457--473.

\bibitem{goyal2006attribute}
V.~Goyal, O.~Pandey, A.~Sahai, and B.~Waters, ``Attribute-based encryption for
  fine-grained access control of encrypted data,'' in \emph{Proceedings of ACM
  CCS}, 2006.

\bibitem{guo2012paas}
L.~Guo, C.~Zhang, J.~Sun, and Y.~Fang, ``Paas: A privacy-preserving
  attribute-based authentication system for ehealth networks,'' in
  \emph{Proceedings of IEEE ICDCS}, 2012.

\bibitem{waters2011ciphertext}
B.~Waters, ``Ciphertext-policy attribute-based encryption: An expressive,
  efficient, and provably secure realization,'' in \emph{PKC}.\hskip 1em plus
  0.5em minus 0.4em\relax Springer, 2011, pp. 53--70.

\bibitem{simoens2013scalable}
P.~Simoens, Y.~Xiao, P.~Pillai, Z.~Chen, K.~Ha, and M.~Satyanarayanan,
  ``Scalable crowd-sourcing of video from mobile devices,'' in
  \emph{Mobisys}.\hskip 1em plus 0.5em minus 0.4em\relax ACM, 2013.

\bibitem{reza2004sse}
R.~Curtmola, J.~Garay, S.~Kamara, and R.~Ostrovsky, ``Searchable symmetric
  encryption: improved definitions and efficient constructions,'' in
  \emph{CCS}.\hskip 1em plus 0.5em minus 0.4em\relax ACM, 2006.

\bibitem{chang2005privacy}
Y.-C. Chang and M.~Mitzenmacher, ``Privacy preserving keyword searches on
  remote encrypted data,'' in \emph{Applied Cryptography and Network
  Security}.\hskip 1em plus 0.5em minus 0.4em\relax Springer, 2005, pp.
  442--455.

\bibitem{wang2012enabling}
C.~Wang, N.~Cao, K.~Ren, and W.~Lou, ``Enabling secure and efficient ranked
  keyword search over outsourced cloud data,'' \emph{TPDS}, vol.~23, no.~8, pp.
  1467--1479, 2012.

\bibitem{li2012toward}
M.~Li, S.~Yu, W.~Lou, and Y.~T. Hou, ``Toward privacy-assured cloud data
  services with flexible search functionalities,'' in \emph{ICDCSW}.\hskip 1em
  plus 0.5em minus 0.4em\relax IEEE, 2012.

\bibitem{wang2012achieving}
C.~Wang, K.~Ren, S.~Yu, and K.~M.~R. Urs, ``Achieving usable and
  privacy-assured similarity search over outsourced cloud data,'' in
  \emph{INFOCOM}.\hskip 1em plus 0.5em minus 0.4em\relax IEEE, 2012, pp.
  451--459.

\bibitem{renprivacy}
Y.~Ren, Y.~Chen, J.~Yang, and B.~Xie, ``Privacy-preserving ranked multi-keyword
  search leveraging polynomial function in cloud computing,'' in
  \emph{Globecom}.\hskip 1em plus 0.5em minus 0.4em\relax IEEE, 2014.

\bibitem{sun2014protecting}
W.~Sun, S.~Yu, W.~Lou, Y.~T. Hou, and H.~Li, ``Protecting your right:
  Attribute-based keyword search with fine-grained owner-enforced search
  authorization in the cloud,'' in \emph{IEEE INFOCOM}, 2014.

\bibitem{chen2012large}
Y.~Chen, B.~Peng, X.~Wang, and H.~Tang, ``Large-scale privacy-preserving
  mapping of human genomic sequences on hybrid clouds.'' in \emph{NDSS}, 2012.

\end{thebibliography}
